\documentclass{fundam}

\publyear{25}
\papernumber{4}
\volume{194}
\issue{4}
\theDOI{10.46298/fi.14425}

\versionForARXIV

\usepackage{url} % takes care of hyperlinks, preferred over hyperref
\usepackage[ruled,lined]{algorithm2e}% provides Algorithm environment
\usepackage{graphicx}% allows for inclusion of graphic files (figures)
\usepackage{tikz}
\usetikzlibrary{automata, positioning, arrows}

\usepackage{amsfonts}
\usepackage{amssymb}
\usepackage{framed}
\usepackage{listings}
\usepackage{makecell}
\usepackage{ctable}
\usepackage{amsmath}
\usepackage[
n,
operators,
advantage,
sets,
adversary,
landau,
probability,
notions,	
logic,
ff,
mm,
primitives,
events,
complexity,
asymptotics,
keys]{cryptocode}

\usepackage{mathrsfs}

\newcommand{\vr}[1]{\mathbf{#1}}
\newcommand{\Lat}{\mathcal{L}}

\newcommand{\Exp}[2]{\mathbf{Exp}_{#1}^{#2}}
\newcommand{\Next}{\mathcal{O}.\mathsf{Next}}
\newcommand{\Upd}{\mathcal{O}.\mathsf{Upd}}
\newcommand{\UpdC}{\mathcal{O}.\mathsf{Upd}\widetilde{\mathsf{C}}}
\newcommand{\Corr}{\mathcal{O}.\mathsf{Corr}}
\newcommand{\Chall}{\mathcal{O}.\mathsf{Chall}}
\newcommand{\OEnc}{\mathcal{O}.\mathsf{Enc}}
\newcommand{\ODec}{\mathcal{O}.\mathsf{Dec}}
\newcommand{\Set}[1]{\mathcal{#1}}
\newcommand{\opuni}{\mathsf{op}\md\mathsf{uni}}
\newcommand{\funi}{\mathsf{f}\md\mathsf{uni}}
\newcommand{\uekg}{\mathsf{UE.KG}}
\newcommand{\uetg}{\mathsf{UE.TG}}
\newcommand{\ueenc}{\mathsf{UE.Enc}}
\newcommand{\uedec}{\mathsf{UE.Dec}}
\newcommand{\ueupd}{\mathsf{UE.Upd}}
\newcommand{\UE}{\mathsf{UE}}
\newcommand{\LWE}{\mathsf{LWE}}
\newcommand{\EUmod}[1]{\mathsf{rand}\md\mathsf{ind}\md\mathsf{eu}\md\mathsf{cpa}\md{#1}}
\newcommand{\EUS}{\mathsf{rand}\md\mathsf{ind}\md\mathsf{eu}\md\mathsf{cpa}}
\renewcommand{\indcpa}{\mathsf{ind}\md\mathsf{cpa}}
\newcommand{\Adv}{\mathsf{Adv}}

\newcommand{\params}{\mathsf{parameters}}

\newcommand{\md}{{\text{-}}}

\newcommand{\ct}{\mathsf{ct}}
\newcommand{\qid}{\mathsf{qid}}
\newcommand{\m}{\mathbf{m}}
\newcommand{\epoch}{\mathsf{e}}
\newcommand{\Epoch}{\mathsf{E}}
\newcommand{\tok}{\Delta}
\newcommand{\tx}[1]{\mathsf{#1}}
\renewcommand{\opuni}{\mathsf{b}\md\mathsf{uni}}

\renewcommand{\Lat}{\mathcal{L}}
\newcommand{\Fun}{\mathcal{F}}
\newcommand{\Par}{\mathcal{P}}
\newcommand{\vol}{\text{vol}}

\newcommand{\Distr}{\mathcal{D}}

\newtheorem{concl}[definition]{Conclusion}
\renewcommand{\sample}{\overset{\$}{\gets}}

\newcommand\quotient[2]{
	\mathchoice
	{% \displaystyle
		\hspace{0.1cm}\text{\raise1ex\hbox{$#1$}\Big/\lower1ex\hbox{$#2$}}
	}
	{% \textstyle
		#1\,/\,#2
	}
	{% \scriptstyle
		#1\,/\,#2
	}
	{% \scriptscriptstyle  
		#1\,/\,#2
	}
}
\begin{document}

\title{Unidirectional Key Update in Updatable Encryption, Revisited}

\address{Military University of Technology, Kaliskiego 2 00-908 Warsaw}

\author{Mariusz Jurkiewicz, Kamila Prabucka\thanks{The authors are grateful to the anonymous reviewers for their thorough reading of the manuscript and for their insightful and constructive feedback.}\thanks{Supported by the Grant DOB/002/RON/ID1/2018 by The National Centre for Research and Development.}\\
	Military University of Technology, Warsaw, Poland\\
	\{mariusz.jurkiewicz, kamila.prabucka\}@wat.edu.pl}

\maketitle

\runninghead{M. Jurkiewicz, K. Prabucka}{Unidirectional Key Update in Updatable Encryption, Revisited}

\begin{abstract}
	We introduce a new, efficient updatable encryption ($\UE$) scheme based on $\tx{FrodoPKE}$, a key encapsulation mechanism grounded in the Learning With Errors ($\LWE$) problem and derived from Frodo KEM, a third-round alternate candidate in the NIST Post-Quantum Cryptography Standardization process. %~\cite{Frodo,Frodo2}. 
The proposed scheme supports uni-directional key updates with backward-leak protection and is proven secure in the $\mathsf{rand}$-$\mathsf{ind}$-$\mathsf{eu}$-$\mathsf{cpa}$ model. Its security relies on the hardness of the $\LWE$ problem, providing strong guarantees against both classical and quantum adversaries under the assumption that $\LWE$ is computationally intractable.
\end{abstract}


\section{Introduction}

Updatable encryption ($\UE$) enables a third party to periodically rotate encryption keys across epochs without requiring the client to download, decrypt, re-encrypt, and re-upload the encrypted data. Instead, ciphertexts can be updated efficiently using \emph{update tokens}, which are compact transformation artifacts generated during key rotation. These tokens can be sent publicly to the storage provider, who applies them to the ciphertexts to bring them in line with the current epoch key, all without learning anything about the underlying plaintexts.

$\UE$ has become an important cryptographic primitive for managing long-term encrypted data in outsourced storage systems, where frequent key rotations are essential for maintaining forward secrecy and minimizing exposure in case of key compromise. Efficient key updates that do not require client-side processing or data access are particularly attractive in cloud-based or distributed settings, where bandwidth, latency, and trust boundaries are significant concerns.

To convey the underlying intuition of $\UE$ while maintaining mathematical rigor, consider a connected and closed set $\Epoch \subseteq \RR_{\geq 0}$ containing $0$, where $\RR_{\geq 0}$ denotes the time axis and $0$ serves as the origin. The set $\Epoch$ is partitioned into a family $\mathcal{E} := \{\epoch_i\}_{i \in \ZZ_{\geq 0}}$ of closed and bounded subsets. If $\Epoch$ is bounded, there exists $i_{\max} \in \ZZ_{\geq 1}$ such that $\epoch_i \neq \emptyset$ for all $i \leq i_{\max}$ and $\epoch_i = \emptyset$ for $i > i_{\max}$. If $\Epoch$ is unbounded, then all sets $\epoch_i$ are nonempty. Furthermore, $\Epoch = \bigcup_{i \in \ZZ_{\geq 0}} \epoch_i$, and for each $i \in \ZZ_{\geq 1}$ with $\epoch_i \neq \emptyset$, we have $\epoch_{i-1} \cap \epoch_i \neq \emptyset$ and $\operatorname{int}(\epoch_{i-1}) \cap \operatorname{int}(\epoch_i) = \emptyset$. Each nonempty set $\epoch_i$ is referred to, in the formal sense, as an \emph{epoch}. It is worth noting that adjacent epochs may share a single point of intersection; this property is convenient for ensuring consistency in the construction and, since such intersections have Lebesgue measure zero on the real line, they do not affect the analysis. We introduce a natural ordering on epochs as follows: we write $\epoch_1 \leq \epoch_2$ if every element of $\epoch_1$ is less than or equal to every element of $\epoch_2$. To emphasize strict ordering, we write $\epoch_1 < \epoch_2$ whenever $\epoch_1 \leq \epoch_2$ and $\epoch_1 \neq \epoch_2$. Moreover, for any epoch $\epoch$, we define its adjacent successor, denoted $\epoch+1$, as the unique epoch with index one greater than that of $\epoch$; similarly, $\epoch+2$ denotes the epoch following $\epoch+1$, and so forth. For convenience, and to avoid unnecessary proliferation of notation, we will also identify each epoch with its index; for example, $\epoch=0$ means that we are actually working with $\epoch_0$. An \textit{updatable encryption scheme} over a nonempty message space $\Set{M}$ consists of a collection of probabilistic polynomial-time (PPT) algorithms $(\uekg, \uetg, \ueenc, \uedec, \ueupd)$ operating over epochs, starting at $\epoch = 0$:
\begin{itemize}
	\item $\uekg$ generates an epoch key $\key_{\epoch}$,
	\item $\uetg$ takes two consecutive epoch keys $\key_{\epoch}$ and $\key_{\epoch+1}$ and outputs an update token $\tok_{\epoch+1}$,
	\item $\ueenc$ encrypts a message $\m \in \Set{M}$ under $\key_{\epoch}$ to produce a ciphertext $\ct_{\epoch}$,
	\item $\uedec$ decrypts a ciphertext $\ct_{\epoch}$ using $\key_{\epoch}$ to recover the message $\m$,
	\item $\ueupd$ takes a token $\tok_{\epoch+1}$ and a ciphertext $\ct_{\epoch}$, and outputs an updated ciphertext for epoch $\epoch+1$.
\end{itemize}

We say that a $\UE$ scheme is \emph{correct} if, for any message $\m \in \Set{M}$ and any two epochs $\epoch_1, \epoch_2 \in \mathcal{E}$ with $\epoch_1 < \epoch_2$, a ciphertext encrypted at epoch $\epoch_1$ and sequentially updated through each intermediate epoch up to $\epoch_2$ can be correctly decrypted under the final key, except with negligible probability. Formally, correctness requires that
\[
\Pr\left[ \uedec(\key_{\epoch_2}, \ct_{\epoch_2}) \neq \m \right] \leq \negl.
\]
In this process, the keys $\key_{\epoch_1}, \ldots, \key_{\epoch_2}$ are generated using $\uekg(1^n)$. The message $\m$ is encrypted under $\key_{\epoch_1}$ to produce the initial ciphertext $\ct_{\epoch_1} \gets \ueenc(\key_{\epoch_1}, \m)$, and for each $j \in \{\epoch_1 + 1, \ldots, \epoch_2\}$, an update token $\tok_j \gets \uetg(\key_{j-1}, \key_j)$ is generated and applied to update the ciphertext via $\ct_j \gets \ueupd(\tok_j, \ct_{j-1})$.

In this work, we present a new $\UE$ scheme that is secure in the $\mathsf{rand}\md\mathsf{ind}\md\mathsf{eu}\md\mathsf{cpa}$ model and is based on the Learning With Errors ($\LWE$) assumption, a well-established foundation for post-quantum cryptography. Our design builds on $\tx{FrodoPKE}$, a lattice-based key encapsulation mechanism grounded in the $\LWE$ problem and known for its simplicity and conservative assumptions. The resulting scheme is secure against both classical and quantum adversaries and achieves strong directional guarantees: it supports \emph{backward-leak uni-directional updates}, meaning the past key $\key_{\epoch}$ can be derived from the future key $\key_{\epoch+1}$ and the update token $\tok_{\epoch+1}$, but not vice versa. This directionality is significant: as shown in prior work~\cite{nishimaki2022direction}, backward-leak uni-directional schemes offer stronger security than both forward-leak and bi-directional models, establishing a clear hierarchy among $\UE$ constructions.

The security of our scheme against both classical and quantum adversaries does not rely solely on the properties of $\tx{FrodoPKE}$. Rather, it is a consequence of the fact that every element of the construction is carefully designed within the framework of lattice-based cryptography. The scheme employs only well-established operations, including Gaussian noise sampling, matrix-vector multiplication over integer lattices, and arithmetic on lattice-structured data, all of which are known to preserve security in both classical and quantum models. The design explicitly avoids relying on computational assumptions that are known to be vulnerable to quantum attacks, such as those based on factoring or discrete logarithms. Because the construction remains entirely within the domain of assumptions grounded in the hardness of the $\LWE$ problem, it maintains strong security guarantees throughout. This argument is formally supported by Theorem~\ref{th::main::one}, which establishes a reduction from breaking the $\EUS$-security of our scheme in the backward-leak setting to solving the $\LWE$ problem, under standard lattice assumptions.

A key advantage of our construction is its efficiency. Unlike some earlier schemes where ciphertext or key sizes grow with the number of epochs, our scheme maintains constant-size ciphertexts and keys, regardless of how many times updates are performed. This makes it particularly well-suited for real-world applications requiring scalability and long-term security guarantees.

In summary, our construction advances the state of updatable encryption by combining three desirable properties: strong directional security in the backward-leak model, post-quantum resilience via the $\LWE$ assumption, and practical efficiency with constant-size ciphertexts and keys. This positions our scheme as a compelling option for secure, scalable, and future-proof encrypted storage.

\section{Related Work and Discussion}

The concept of updatable encryption was formally introduced by Boneh et al.~\cite{boneh2013key}, marking the beginning of a broader investigation into secure ciphertext updates across key epochs. Early efforts focused on ciphertext-dependent schemes, where each update token is generated relative to a specific ciphertext. This line of work was initiated by Everspaugh et al.~\cite{EPRS17}, and subsequently refined by Boneh et al.~\cite{BEKS20} and Chen et al.~\cite{CLT20}, who expanded the model and proposed stronger security definitions to better capture realistic adversarial capabilities.

A notable advancement was made by Lehmann and Tackmann~\cite{LT18}, who introduced ciphertext-independent $\UE$, allowing the same update token to apply uniformly across all ciphertexts within a given epoch. They also explored bi-directional key updates, where ciphertexts can be updated both forward and backward in time. This framework was refined by follow-up work from Klooss et al.~\cite{klooss2019r}, Boyd et al.~\cite{BDGJ20}, and Jiang~\cite{jiang2020}, who examined the interplay between directionality, update efficiency, and security.

A major theoretical contribution came from Nishimaki~\cite{nishimaki2022direction}, who emphasized the importance of update direction in determining security guarantees. He distinguished between forward-leak and backward-leak uni-directional schemes: forward-leak tokens can derive future keys, while backward-leak tokens derive only past keys. He demonstrated that backward-leak uni-directionality provides strictly stronger security than both forward-leak and bi-directional models, establishing a clear hierarchy among directional schemes. This result redefined the understanding of secure key updates and has since become a foundational insight in $\UE$ research.

Jiang and Pan~\cite{jiang2023backward} extended this line of research by showing that backward-leak uni-directional schemes can achieve security guarantees equivalent to those of no-directional schemes, which currently represent the strongest known model. Independently, Slamanig and Striecks~\cite{SS21} proposed a pairing-based, no-directional $\UE$ construction that incorporates an expiry mechanism. In their scheme, each ciphertext is associated with an explicit expiration epoch $e_\perp$, beyond which it becomes undecryptable. Specifically, if a token $\Delta_{e+1}$ is used to update a ciphertext past its expiry (i.e., when $e + 1 > e_\perp$), the ciphertext becomes irrecoverable. This enhanced model invalidates the equivalence result previously established by Jiang~\cite{jiang2020} and enables uni-directional updates in a setting that does not impose directional constraints. Their construction is proven secure under the SXDH assumption, with ciphertext and key sizes growing as $\bigO{(\log^2 T)}$, where $T$ denotes the total number of epochs. In contrast, Nishimaki~\cite{nishimaki2022direction} achieves post-quantum security in the backward-leak model while keeping ciphertext and key sizes independent of $T$. His scheme builds on a variant of the Regev encryption scheme~\cite{regev2009lattices}, leveraging the hardness of the $\LWE$ problem to ensure resistance against quantum attacks. Our construction follows this approach: it also achieves post-quantum security in the backward-leak uni-directional setting and maintains constant ciphertext and key sizes, regardless of the number of epochs.

%The construction proposed in this paper is based on the computational hardness of the Learning With Errors (LWE) problem, introduced by Regev~\cite{regev2009lattices}, which is widely believed to be intractable even for quantum computers. This assumption forms the core of the scheme’s security. Specifically, our approach employs FrodoPKE~\cite{Frodo,Frodo2}, a lattice-based public-key encryption scheme whose security directly relies on the LWE assumption~\cite{regev2009lattices}. As a result, FrodoPKE provides strong protection against both classical and quantum adversaries, making it well-suited for post-quantum cryptographic applications. By combining the directional security guarantees of backward-leak updatable encryption~\cite{nishimaki2022direction} with the proven quantum resistance of FrodoPKE~\cite{Frodo,Frodo2}, our construction achieves an effective balance between efficiency and robustness. Compared to earlier bi-directional and pairing-based approaches~\cite{LT18, BDGJ20}, it delivers notable improvements in both practical deployment and long-term security.

The construction proposed in this paper is based on the computational hardness of the $\LWE$ problem, introduced by Regev~\cite{regev2009lattices}, which is widely believed to be resistant even to quantum attacks. This assumption forms the foundation of our scheme’s security. We specifically build on $\tx{FrodoPKE}$~\cite{Frodo,Frodo2}, a lattice-based public-key encryption scheme whose security is directly grounded in the $\LWE$ assumption~\cite{regev2009lattices}. As a result, $\tx{FrodoPKE}$ offers strong protection against both classical and quantum adversaries, making it highly suitable for post-quantum cryptographic applications. Although $\tx{FrodoPKE}$ and Regev’s original encryption scheme differ in formal aspects such as message encoding, matrix dimensions, and noise generation, they share a common conceptual framework. Both follow the same high-level structure: a public key formed from noisy linear combinations, encryption using randomized inputs and controlled noise, and decryption through careful noise cancellation. $\tx{FrodoPKE}$ can therefore be viewed as a robust and scalable realization of Regev’s foundational idea. Given this close conceptual alignment, it is reasonable to expect that $\tx{FrodoPKE}$ retains the same desirable characteristics as Regev’s scheme when used as a foundation for building $\LWE$-based updatable encryption. By combining the strong directional guarantees of backward-leak updatable encryption~\cite{nishimaki2022direction} with the post-quantum security of $\tx{FrodoPKE}$, our construction achieves an effective balance between efficiency, simplicity, and long-term robustness. Compared to earlier bi-directional and pairing-based approaches~\cite{LT18, BDGJ20}, it offers clear advantages in both practical deployment and future-proof security.

\section{Preliminaries}
For a positive integer $k$, let $[k]:=\{1,\ldots k\}$, and $[k]_{0}:=\{0,1,\ldots k\}$. Denote $\ell_{2}$ and $\ell_{\infty}$ norm by $\|\cdot\|$ and $\|\cdot\|_{\infty}$, respectively. 

Vectors are in column form and they are denoted by bold lower case letters (e.g., $\vr{x}$). We view a matrix as the set of its column vectors and denote by bold capital letters (e.g., $\vr{A}$). The $i$th entry of a vector $\vr{x}$ is denoted $x_{i}$, and the $j$th column of a matrix $\vr{A}$ is denoted $\vr{a}_{j}$ or $\vr{A}[j]$. We identify a matrix $\vr{A}$ with the ordered set $\{\vr{a}_{j}\}$ of its column vector. For convenience, we define the norm of a matrix $\vr{A}$ as the maximum absolute value of its entries, i.e. $\|\vr{A}\|_{\max} = \max_{j}\|\vr{a}_{j}\|_{\infty}$. Note that usually $\|\vr{A}\|_{\max} \neq \|\vr{A} \|_{\infty} $ ($=\sup_{\|\vr{x}\|_{\infty}=1}\|\vr{A}\vr{x}\|_{\infty}$). Roughly speaking, $\|\cdot \|_{\max}$ treats $\vr{A}$ more as a (multi) vector rather than an operator. If the columns of $\vr{A} = \{ \vr{a}_{1}, \ldots ,  \vr{a}_{k} \}$ are linearly independent, then $\vr{A}^{*} = \{ \vr{a}_{1}^{*}, \ldots ,  \vr{a}_{k}^{*} \} $ denote the Gram-Schmidt orthogonalization of vectors $\vr{a}_{1}, \ldots ,  \vr{a}_{k}$ taken in that order. For $\vr{A}\in\RR^{n\times m_{1}}$ and $\vr{B}\in\RR^{n\times m_{2}}$, having an equal number of rows, $[\vr{A}|\vr{B}]\in\RR^{n\times (m_{1}+m_{2})}$ denotes the concatenation of the columns of $\vr{A}$ followed by the columns of $\vr{B}$. Likewise, for  $\vr{A}\in\RR^{n_{1}\times m}$ and $\vr{B}\in\RR^{n_{2}\times m}$, having an equal number of columns, $[\vr{A};\vr{B}]\in\RR^{(n_{1}+n_{2})\times m}$ is the concatenation of the rows of $\vr{A}$ and the rows of $\vr{B}$.

Let $I$ be a countable set, and let $\Set{X}=\{X_{n}\}_{n\in I}$, $\Set{Y}=\{Y_{n}\}_{n\in I}$ be two families of random variables such that $X_{n}, Y_{n}$ take values in a finite set $\mathcal{R}_{n}$. We call $\Set{X}$ and $\Set{Y}$ \textit{statistically/perfectly indistinguishable} if  their statistical distance is negligible, where the statistical distance between $\{X_{n}\}_{n\in I}$ and $\{Y_{n}\}_{n\in I}$ is defined as the function
%\begin{align*}
	$\Delta(X_{n}, Y_{n}) = \frac{1}{2}\sum_{r\in \mathcal{R}_{n}} \left| \Pr[X_{n}=r] - \Pr[Y_{n}=r]  \right|.$
%\end{align*}
(See \cite{goldreich2001foundations} for more details).

\begin{lemma}[smudging lemma {\normalfont(\cite{asharov2012multiparty})}]\label{lem::smudging}
	Let $B_{1}=B_{1}(n)$, and $B_{2}=B_{2}(n)$ be positive integers and let $e_{1}\in [-B_{1}, B_{1}]$ be a fixed integer. Let $e_{2}\sample [-B_{2}, B_{2}]$ be chosen uniformly at random. If $B_{1}/B_{2}=\negl$, the distribution of $e_{2}$ is statistically indistinguishable from that of $e_{2}+e_{1}$.
\end{lemma}

\subsection{Lattices}\label{Lattices}
In this subsection, we define lattices and review some of their fundamental properties. For a thorough and accessible introduction to lattice theory, we refer the reader to~\cite{peikert2015decade}.

\vspace{0.5cm}
%Any discrete additive subgroup $\Lat$ of $\RR^{m}$ is called an $m$-dimensional \textit{lattice}, i.e. $\Lat$ is a lattice iff the following conditions hold:
%\begin{enumerate}
%	\item $\vr{0}\in \Lat$, and if $\vr{v}, \vr{w}\in\Lat$ then $-\vr{v}, \vr{v}+\vr{w}\in \Lat$; 
%	\item there is $\varepsilon >0$ such that for every $\vr{v}\in \Lat$ we have $\Lat\cap \{ \vr{w}\mid \| \vr{v}-\vr{w}\|<\varepsilon \} = \{\vr{v}\}$.
%\end{enumerate}
A subset $\Lat \subseteq \RR^m$ is called an $m$-dimensional \textit{lattice} if it is a discrete additive subgroup of $\RR^m$. That is, $\Lat$ is a lattice if and only if the following conditions are satisfied:
\begin{enumerate}
	\item $\vr{0} \in \Lat$, and for any $\vr{v}, \vr{w} \in \Lat$, we have $-\vr{v},\, \vr{v} + \vr{w} \in \Lat$;
	\item there exists $\varepsilon > 0$ such that for every $\vr{v} \in \Lat$, the intersection $\Lat \cap \{ \vr{w} \in \RR^m \mid \| \vr{v} - \vr{w} \| < \varepsilon \}$ contains only $\vr{v}$.
\end{enumerate}
Because a lattice $\Lat$ is is an additive subgroup of $\RR^{m}$, it induces the quotient group $\quotient{\RR^{m}}{\Lat}$ of cosets
\begin{align*}
	\vr{x} + \Lat = \{ \vr{x}+\vr{v} \mid \vr{v}\in\Lat  \}, ~~~\vr{x}\in\RR^{m},
\end{align*}
with respect to the addition operation $(\vr{x}+\Lat) + (\vr{y}+\Lat) = (\vr{x}+\vr{y}) + \Lat$.

Lattices admit a particularly clear representation: for any lattice $\Lat$, there exists a set of linearly independent vectors $\vr{B} = \{\vr{b}_1, \ldots, \vr{b}_k\}$ such that every point in $\Lat$ is an integer linear combination of the vectors in $\vr{B}$. In other words,
\begin{align*}
	\Lat = \Lat(\vr{B}) = \vr{B} \cdot \ZZ^k = \left\{ \sum_{i=1}^{k} \alpha_i \vr{b}_i \mid \alpha_i \in \ZZ \right\}.
\end{align*}
%It turns out that lattices have clearer representation, namely for a lattice $\Lat$ there is a set $\vr{B}$  consisting of linearly independent vectors $\vr{B} = \{\vr{b}_{1}, \ldots , \vr{b}_{k}\}$ such that any lattice point is a integer linear combinations of vectors form $\vr{B}$. 
%\begin{align*}
%	\Lat = \Lat (\vr{B})=\vr{B}\cdot \ZZ^{k} = \left\{  \sum_{i=1}^{k}\alpha_{i}\vr{b}_{i}\mid \alpha_{i}\in\ZZ  \right\}.
%\end{align*}
The set $\vr{B}$ is called a \textit{basis} of the lattice $\Lat$, and its cardinality $k = \#\vr{B}$ is referred to as the \textit{rank} of $\Lat$.  
If $k = m$, we say that $\Lat$ is a \textit{full-rank} lattice.  
A lattice basis $\vr{B}$ is not unique; in fact, for any unimodular matrix $\vr{U} \in \ZZ^{m \times m}$ (i.e., a matrix with $|\det(\vr{U})| = 1$), the set $\vr{B} \cdot \vr{U}$ is also a basis of $\Lat(\vr{B})$.

A \emph{fundamental domain} of $\Lat$ is a connected set $\Fun\subset \RR^{m}$ such that $\vr{0}\in\Fun$ and it contains exactly one representative $\bar{\vr{x}}$ of every coset $\vr{x}+\Lat$. For a lattice $\Lat$ having basis $\vr{B}$, a commonly used fundamental domain is the origin-centered fundamental parallelepiped $\Par(\vr{B}) = \vr{B}\cdot \left( -\frac{1}{2}, \frac{1}{2}  \right]^{m}$, where a coset $\vr{x}+\Lat$ has representative $\vr{x} - \vr{B}\cdot \lfloor \vr{B}^{-1}\cdot \vr{x}\rceil$. The measure of fundamental parallelepiped  can be easily computed as $\vol \Par(\vr{B}) = \sqrt{\det\vr{B}\cdot \vr{B}^{T}}$, in addition this number does not depend on the choice of bases of the lattice, i.e. if $\Lat = \Lat(\vr{B}_{1}) = \Lat(\vr{B}_{2})$ then $\vol \Par(\vr{B}_{1}) = \vol \Par(\vr{B}_{2})$. Therefore, the measure of fundamental parallelepipeds is invariant of the lattice and is called the \textit{determinant} of the lattice $\Lat$ and denoted by $\det\Lat$.

A full-rank lattice $\Lat$ is called an \textit{integer lattice} if $\Lat \subseteq \ZZ^m$. An integer lattice is called a \textit{$q$-ary lattice} if it satisfies $q\ZZ^m \subseteq \Lat \subseteq \ZZ^m$ for some $q \in \ZZ_{\geq 1}$.

By definition, a lattice $\Lat = \Lat(\vr{B})$ is an integer lattice if and only if $\vr{B} \in \ZZ^{m \times m}$ is an integer square matrix. In this case, the determinant $\det \Lat = \abs{\det{\vr{B}}}$ is a positive integer. It is easy to show that every integer lattice is also a $(\det \Lat)$-ary lattice. 
For a $q$-ary lattice $\Lat$, the following elementary algebraic facts hold:
\begin{itemize}
	\item $\ZZ^{m} / q\ZZ^{m} \cong \ZZ_q^m$ (as additive groups);
	\item $\ZZ^{m} / \Lat \cong \ZZ_q^m / (\Lat \bmod q)$ (as additive groups).
\end{itemize}
Moreover, $\ZZ^{m} / \Lat$ is a finite group, and its order is given by $|\ZZ^{m} / \Lat| = \det \Lat$.

\subsection{Gaussians}\label{Gaussian}
For any real $s>0$ and $\vr{c}\in\RR^{m}$, the \emph{Gauss function} $\rho_{s,\vr{c}}$ centered on $\vr{c}$ with parameter $s$ is defined as
\begin{align*}
	\rho_{s,\vr{c}}(\vr{x})  = \exp \left(-\frac{\pi}{s^{2}}\|\vr{x} - \vr{c} \|^{2}\right), ~~~\vr{x}\in\RR^{m},
\end{align*}
and 
\begin{align*}
	\rho_{s} = \rho_{s, \vr{0}}, ~~~\rho = \rho_{1},~\text{i.e.}~ \rho(\vr{x}) = e^{-\pi\| \vr{x} \|^{2}}.  
\end{align*}
%Directly from the definition we have $\rho_{s}(\vr{x}) = \rho\left(  {s}^{-1} \vr{x}\right)$.
By definition, we have $\rho_s(\mathbf{x}) = \rho(s^{-1} \mathbf{x})$.

For any real $s>0$, \textit{the rounded Gaussian distribution} with parameter (or width) $s$, denoted $\Psi_{s}$, is the distribution over $\ZZ$ obtained by rounding a sample from $\Distr_{s}$ to the nearest integer:
\begin{align*}
	\Psi_{s}(x)= \int_{\{ z\mid \lfloor z \rceil = x\}} \Distr_{s}(z)dz.
\end{align*}
Let $\Lat\subseteq \ZZ^{m}$ be a lattice and let $\rho_{s, \vr{c}} (\Lat) = \sum_{\vr{x}\in \Lat} \rho_{s, \vr{c}}(\vr{x})$. Define the \emph{discrete Gaussian distribution} over $\Lat$ with center $\vr{c}$, and parameter $s$ as
\begin{align*}
	\Distr_{\Lat, s, \vr{c}} (\vr{x}) =  \frac{\rho_{s, \vr{c}}(\vr{x})}{\rho_{s, \vr{c}}(\Lat)}, ~~~\vr{x}\in\Lat.
\end{align*}
For notational convenience, we let
\begin{align*}
	\Distr_{\Lat, s} = \Distr_{\Lat, s, \vr{0}}, ~ \Distr_{s, \vr{c}}^{m} = \Distr_{\ZZ^{m}, s, \vr{c}}, ~ \Distr_{s}^{m} = \Distr_{\ZZ^{m}, s}.
\end{align*}

%\begin{lemma}{\normalfont (\cite{banaszczyk1993new,lindner2011better})}
%	For any $s,t\in\RR_{>0}$, $c\in\RR_{\geq1}$, $C=c\cdot \exp{\left( 2^{-1}\cdot(1-c^{2}) \right)}$, $m\in\ZZ_{>0}$, and any $\vr{y}\in\RR^{m}$ the following conditions hold: 
%	\begin{itemize}
	%		\item $\Pr_{\vr{x}\gets  \Distr_{s}^{m}} \left[ \| \vr{x}\|_{\infty}> t\cdot s \right]\leq 2e^{-\pi t^{2}}$;
	%		\item $\Pr_{\vr{x}\gets  \Distr_{s}^{m}} \left[ \| \vr{x}\|_{\infty}>   c\cdot \frac{1}{\sqrt{2\pi}} \cdot s\sqrt{m}   \right]\leq C^{m}$;
	%		\item $\Pr_{\vr{x}\gets  \Distr_{s}^{m}} \left[    \vert  \inner{x}{y} \vert  > t\cdot s\norm{y} \right]         \leq 2e^{-\pi t^{2}}$.
	%	\end{itemize}
%\end{lemma}

\bigskip

We collect several known results from the literature concerning discrete Gaussians over lattices, specialized to the family relevant to our setting.
\begin{lemma}\label{lem::02}
	Let $n < m$, and let $\mathbf{T}$ be any basis of the lattice $\Lat$ corresponding to some matrix $\mathbf{A} \in \mathbb{Z}_q^{n \times m}$ whose columns generate $\mathbb{Z}_q^n$. Let $\mathbf{u} \in \mathbb{Z}_q^n$ and $\mathbf{c} \in \mathbb{Z}^m$ be arbitrary, and assume that $s \geq \| \mathbf{T}^* \| \cdot \omega\left( \sqrt{\log m} \right)$. Then:
	
	\begin{enumerate}
		\item {\normalfont \cite{micciancio2007worst,cash2012bonsai}:} 
		$\Pr_{\mathbf{x} \gets \mathcal{D}_{\Lat, s, \mathbf{c}}} \left[ \| \mathbf{x} - \mathbf{c} \| > s \cdot \sqrt{m} \right] \leq \negl.$
		\item {\normalfont \cite{peikert2006efficient,cash2012bonsai}:} 
		$\Pr_{\mathbf{x} \gets \mathcal{D}_{\Lat, s}} \left[ \mathbf{x} = 0 \right] \leq \negl.$
	\end{enumerate}
\end{lemma}

\section{Security Model for Updatable Encryption}
In this section, we present the formal security model for updatable encryption. We begin by defining the leakage sets and then describe key leakage and token leakage, followed by a detailed description of the adversary’s available oracles within the security game. Finally, we outline the indistinguishability experiment under chosen-plaintext attacks, which forms the foundation of our security definition.

\vspace{0.3cm}
\noindent\textbf{Leakage sets.} The security notions for $\mathsf{UE}$ are based on the concept of \textit{leakage sets}, originally introduced by Klooss et al.~\cite{klooss2019r} and Jiang~\cite{jiang2020}. These sets are used to track the epochs in which the adversary has compromised a key, an update token, or a version of a challenge-equal ciphertext.
\begin{itemize}
	\item $\Set{K}$: set of epochs in which the adversary corrupted the epoch key through $\Corr$; 
	\item $\Set{T}$: set of epochs in which the adversary corrupted the update token through $\Corr$; 
	\item $\Set{C}$: set of epochs in which the adversary obtained a challenge-equal ciphertext through $\Chall$ or $\UpdC$.
\end{itemize}
In addition to these sets, it is also necessary to track ciphertexts and their updates that are accessible to the adversary:
\begin{itemize}
	\item $\Set{L}$: the set of non-challenge ciphertexts of the form $(\qid, \ct, \epoch; \m)$, where $\qid$ is a query identifier incremented with each call to $\OEnc$, $\ct$ is the ciphertext, $\epoch$ is the epoch in which the encryption occurred, and $\m$ is the plaintext. These ciphertexts are learned through $\OEnc$ or $\Upd$;
	\item $\widetilde{\Set{L}}$: the set of challenge-equal ciphertexts of the form $(\widetilde{\ct}, \epoch)$, where $\widetilde{\ct}$ is the challenge-equal ciphertext and $\epoch$ is the associated epoch. These ciphertexts are revealed through $\Chall$ or $\UpdC$.
\end{itemize} 
The sets $\Set{L}$ and $\widetilde{\Set{L}}$, along with their respective extensions, are used exclusively in the deterministic variants of $\UE$, as detailed by Jiang~\cite{jiang2020}. In the randomized setting, however, two plaintext-based sets, $\Set{Q}^{*}$ and $\widetilde{\Set{Q}}^{*}$, are introduced instead. Specifically, $\Set{Q}^{*}$ consists of pairs $(\m, \epoch)$ such that the adversary has learned or is able to produce a ciphertext in epoch $\epoch$ under the message $\m$. The set $\widetilde{\Set{Q}}^{*}$ consists of pairs $(\bar{\m}, \epoch)$ or $(\bar{\m}_1, \epoch)$, where $(\bar{\m}, \bar{\ct})$ is the input to the challenge query $\mathcal{O}.\mathsf{Chall}$ and $\bar{\m}_1$ is the underlying message of the challenge ciphertext $\bar{\ct}$. These represent challenge-equal ciphertexts that the adversary can generate in epoch $\epoch$ under either message. This distinction enables a one-to-one correspondence between ciphertexts and their associated messages in both the deterministic and randomized settings. In contrast, the scheme presented in this paper is purely randomized, and we therefore do not consider the deterministic case.

Note that, since the adversary may learn or infer additional information through known tokens, the extended sets $\Set{K}^{*}$, $\Set{T}^{*}$, $\Set{C}^{*}$, $\Set{L}^{*}$, and $\widetilde{\Set{L}}^{*}$ are defined as supersets of the original sets $\Set{K}$, $\Set{T}$, $\Set{C}$, $\Set{L}$, and $\widetilde{\Set{L}}$, respectively.

\vspace{0.3cm}
\noindent\textbf{Key Leakage.} Key updates can be classified into three distinct cases; see Jiang~\cite{jiang2020} for a more detailed treatment:
\begin{itemize}
	\item the pair $(\key_{\epoch}, \tok_{\epoch+1})$ does not allow the derivation of $\key_{\epoch+1}$ ($\tx{no}$-directional key update);
	\item the pair $(\key_{\epoch}, \tok_{\epoch+1})$ allows the derivation of $\key_{\epoch+1}$ ($\tx{uni}$--directional key update);
	\item either $(\key_{\epoch}, \tok_{\epoch+1})$ or $(\key_{\epoch+1}, \tok_{\epoch+1})$ enables the derivation of $\key_{\epoch+1}$ or $\key_{\epoch}$, respectively ($\tx{bi}$-directional key update).
\end{itemize}

At first glance, it may seem that a $\tx{bi}$-directional key update implies a $\tx{uni}$-directional key update. However, this is not the case, as unidirectionality is defined precisely by the restriction that updates are possible in only one direction. In particular, this property does not hold in the $\tx{bi}$-directional setting.
As noted by Nishimaki~\cite{nishimaki2022direction}, the conventional classification into $\tx{no}$-, $\tx{uni}$-, and $\tx{bi}$-directional key updates does not encompass all possible cases. He highlighted a particularly important additional scenario: a reverse form of unidirectionality, in which knowledge of $(\key_{\epoch+1}, \tok_{\epoch+1})$ enables the derivation of $\key_{\epoch}$. This case merits separate treatment as it represents a distinct class of key update directionality. As a result, the notion of $\tx{uni}$-directional key update was refined into two distinct subtypes: \textit{forward-leak uni-directional key update} ($\funi$), which corresponds to the original $\tx{uni}$ model (i.e., $\Set{K}_{\funi}^{*} = \Set{K}_{\mathsf{uni}}^{*}$), and \textit{backward-leak uni-directional key update} ($\opuni$). If $l+1 \in \ZZ_{>0}$ denotes the total number of epochs, the set $\Set{K}_{\opuni}^{*}$ is defined as follows:

$$\Set{K}_{\opuni}^{*}\gets \set{\epoch\in [0,l]\mid \tx{CorrK}(\epoch)=\tx{true}},$$ $$ \text{ where } \tx{true}\gets\tx{CorrK}(\epoch)\Leftrightarrow (\epoch \in\Set{K}) \lor (\tx{CorrK(\epoch +1 )\land (\epoch +1 )\in \Set{T}}).$$
Interestingly, $\opuni$-directional key updates provide stronger security guarantees than their $\tx{bi}$-directional counterparts. Furthermore, Jiang, Galteland, and Pan~\cite{jiang2023backward} established a striking result: security in the backward-leak uni-directional setting is equivalent to that of the no-directional setting.

\vspace{0.3cm}
\noindent\textbf{Token Leakage.} A token is considered known to the adversary if it has either been explicitly corrupted or can be inferred from two consecutive epoch keys. Accordingly, the extended set $\Set{T}_{\opuni}^{*}$ is defined as a function of $\Set{T}$ and $\Set{K}_{\opuni}^{*}$:
\[
\Set{T}_{\opuni}^{*} \gets \left\{ \epoch \in [0, l] \,\middle|\, (\epoch \in \Set{T}) \lor \left(\epoch \in \Set{K}_{\opuni}^{*} \land (\epoch - 1) \in \Set{K}_{\opuni}^{*} \right) \right\}.
\]

\vspace{0.3cm}
\noindent\textbf{Oracles in security games for updatable encryption}. The security game corresponding to the variant of $\UE$ considered in this work involves the following oracles: $\mathbf{Setup}$, $\OEnc$, $\ODec$, $\Next$, $\Upd$, $\Corr$, $\Chall$, and $\UpdC$.

\vspace{0.5cm}

\begin{pchstack}[center]
	
	\begin{pcvstack}%
		\procedure{$\mathbf{Setup}(1^{n})$}{%
			\pcln \epoch\gets 0\\
			\pcln \key_{\epoch}\sample \uekg(1^{n}), \tok_{\epoch}\gets \perp \\
			\pcln \qid\gets 0, \tx{twf}\gets 0, \tx{phase}\gets 0 \\
			\pcln \Set{L}, \widetilde{\Set{L}}, \Set{C}, \Set{K}, \Set{T}\gets \emptyset }
		
		\pcvspace
		
		\procedure{$\OEnc(\m)$}{%
			\pcln  \qid \gets \qid +1   \\
			\pcln  \ct \sample \ueenc(\key_{\epoch}, \m)  \\
			\pcln  \Set{L} \gets \Set{L}\cup \set{\qid, \ct, \epoch; \m}  \\
			\pcln   \pcreturn \ct }
		
		\pcvspace
		
		\procedure{$\ODec(\ct)$}{%
			\pcln  \m' ~\tx{or}~\perp \gets \uedec(\key_{\epoch}, \ct) \\
			\pcln \pcif (\m', \epoch)\in \widetilde{\Set{Q}}^{*}~\textbf{then}\\
			\pcln ~~~~~\tx{twf}\gets 1\\
			\pcln   \pcreturn \m'~\textbf{or}~\perp }
	\end{pcvstack}
	
	\pchspace
	\hspace{1.5cm}
	
	\begin{pcvstack}%
		
		\procedure{$\Upd(\ct_{\epoch-1})$}{%
			\pcln  \pcif (j, \ct_{\epoch-1}, \epoch-1, \m)\not\in \Set{L}~\textbf{then}\\
			\pcln~~~~~\pcreturn\perp\\
			\pcln \ct_{\epoch}\gets \ueupd(\tok_{\epoch}, \ct_{\epoch-1})\\
			\pcln \Set{L}\gets \Set{L}\cup \set{(j, \ct_{\epoch}, \epoch; \m)}\\
			\pcln   \pcreturn \ct_{\epoch}}	
		
		\pcvspace
		
		\procedure{$\Corr(\tx{inp}, \hat{\epoch})$}{%
			\pcln  \pcif \hat{\epoch}>\epoch~\textbf{then}\\
			\pcln ~~~~~\pcreturn \perp\\
			\pcln\pcif \tx{inp}=\tx{key}~\textbf{then}\\
			\pcln ~~~~~\Set{K}\gets \Set{K}\cup \set{\hat{\epoch}}\\
			\pcln ~~~~~\pcreturn \key_{\hat{\epoch}}\\
			\pcln \pcif \tx{inp}=\tx{token}~\textbf{then}\\
			\pcln ~~~~~\Set{T}\gets \Set{T}\cup \set{\hat{\epoch}}\\
			\pcln ~~~~~\pcreturn \tok_{\hat{\epoch}}}	
			
	\end{pcvstack}
	
\end{pchstack}

\begin{pchstack}[center]
	
	\begin{pcvstack}%
		\procedure{$\Next()$}{%
			\pcln \epoch\gets \epoch +1 \\
			\pcln \key_{\epoch} \sample \uekg(1^{n})\\
			\pcln \tok_{\epoch}\gets \uetg(\key_{\epoch-1}, \key_{\epoch})\\
			\pcln\pcif \tx{phase}=1~\textbf{then}\\
			\pcln ~~~~~\widetilde{\ct}_{\epoch} \gets \ueupd(\tok_{\epoch}, \widetilde{\ct}_{\epoch\md1})\\
			\pcln ~~~~~\Set{C} \gets \Set{C} \cup \set{\epoch}\\
			\pcln ~~~~~\widetilde{\Set{L}} \gets \widetilde{\Set{L}} \cup \set{(\widetilde{\ct}_{\epoch}, \epoch)}}
			
			\pcvspace
			
		\procedure{$\UpdC()$}{%
			\pcln \pcif \tx{phase}\neq 1~\textbf{then}\\
			\pcln ~~~~~\pcreturn \perp\\
			\pcln \pcreturn \widetilde{\ct}_{\epoch}}
	\end{pcvstack}
	
	\pchspace
	\hspace{1.5cm}
	
	\begin{pcvstack}%
		
		\procedure{$\Chall(\bar{\m}, \bar{\ct}, \tx{b})$}{%
			\pcln\pcif  \tx{phase}=1 ~\textbf{then}  \\
			\pcln ~~~~~\textbf{return}~\perp  \\
			\pcln  \tx{phase}\gets 1; \widetilde{\epoch}\gets \epoch  \\
			\pcln\pcif  (\cdot, \bar{\ct}, \widetilde{\epoch}-1, \bar{\m}_{1})\not\in \Set{L} ~\textbf{then}  \\
			\pcln ~~~~~\textbf{return}~\perp  \\
			\pcln\pcif  \tx{b}=0 ~\textbf{then}\\
			\pcln ~~~~~\widetilde{\ct}_{\widetilde{\epoch}} \gets \ueenc(\key_{\widetilde{\epoch}}, \bar{\m}) \\
			\pcln\pcelse \\
			\pcln ~~~~~\widetilde{\ct}_{\widetilde{\epoch}} \gets \ueupd(\tok_{\widetilde{\epoch}}, \bar{\ct}) \\
			\pcln \Set{C}\gets \Set{C}\cup \set{\widetilde{\epoch}}\\
			\pcln \widetilde{\Set{L}} \gets \widetilde{\Set{L}} \cup \set{(\widetilde{\ct}_{\widetilde{\epoch}}, \widetilde{\epoch})}\\
			\pcln   \pcreturn \widetilde{\ct}_{\widetilde{\epoch}}}		
		
	\end{pcvstack}
	
\end{pchstack}

\vspace{0.5cm}

Each of the algorithms and global variables defined above plays a specific role within the security experiment. Several global technical variables are maintained to accurately track the evolution of the system state as the experiment progresses. The unsigned integer~$\qid$ serves as a universal query counter, increasing with each call to the encryption oracle~($\OEnc$) and uniquely indexing every query instance. Two Boolean variables further regulate the experiment. The variable~$\tx{twf}$ is set to~$1$ whenever a trivial or forbidden event arises; for example, this occurs if the adversary obtains the secret key for a particular epoch, which would enable direct decryption of the challenge ciphertext or any of its updated variants within that epoch. The variable~$\tx{phase}$ indicates the current stage of the experiment, with $\tx{phase} = 1$ indicating entry into the challenge phase. The main algorithms $\mathbf{Setup}$, $\OEnc$, and $\Upd$ adhere to the definition of $\UE$ schemes and function as expected in this context. The decryption oracle~($\ODec$) does not participate in our security proofs, since our analysis is limited to the chosen plaintext attack~($\tx{CPA}$) setting. Its presence serves only for completeness and for consistency with established literature, where broader models such as chosen ciphertext attacks~($\tx{CCA}$) are considered. The $\Next$ oracle advances the experiment to the next epoch by generating a new epoch key and update token. If the experiment is currently in the challenge phase, $\Next$ also updates the challenge ciphertext and records each new version so that all such versions remain available to the adversary for later queries. The $\UpdC$ oracle allows the adversary to obtain the current, updated version of the challenge ciphertext for the present epoch, referencing the state maintained by $\Next$. The $\Corr$ oracle provides a means for the adversary to corrupt keys and tokens, thereby obtaining secret keys or update tokens for epochs of its choice. The $\Chall$ oracle initiates the challenge phase by returning either a freshly encrypted message or an updated ciphertext as the challenge, which commences the central distinguishing experiment. Together, these variables and oracles, combined with the definition of the relevant leakage sets, establish a rigorous and comprehensive framework for formalizing the adversary’s capabilities and for modeling the progression of the security experiment in the context of updatable encryption.

\vspace{0.3cm}
\noindent\textbf{Challenge-Equal Ciphertext Leakage.} The adversary learns all challenge-equal ciphertexts in epochs from the set $\Set{C}$. Additionally, such ciphertexts can be inferred via known update tokens, which may reveal further information through indirect derivations.
It is easy to see that the structure of the extended set $\Set{C}^{*}$ depends on whether ciphertexts are updated in a $\tx{uni}$- or $\tx{bi}$-directional manner. Importantly, it remains independent of whether key and token updates follow the $\opuni$-directional model. Let $\tx{cc} \in \{\tx{uni}, \tx{bi}\}$; the procedure for computing the sets $\Set{C}_{\opuni, \tx{cc}}^{*}$ is described by Jiang~\cite{jiang2020}.

\vspace{0.3cm}
\noindent\textbf{Security of $\UE$ scheme in $\tx{CPA}$ model.}

\vspace{0.5cm}
\begin{pchstack}[center]
	\procedure{$\Exp{\UE,\adv}{\EUmod{0}}$}{%
		\pcln\textbf{do} ~\mathbf{Setup}(1^{n})\\
		\pcln \tx{b}'\gets \adv^{\tx{oracles}(\tx{b}=0)}(1^{n})\\
		\pcln\pcif \Set{K}^{*}\cap \Set{C}^{*}\neq \emptyset~\textbf{then}\\
		\pcln~~~~~\tx{twf}\gets 1\\
		\pcln\pcif \tx{twf}= 1~\textbf{then}\\
		\pcln ~~~~~\tx{b}'\sample \bin\\
		\pcln   \pcreturn \tx{b}' }		
	
	\pchspace
	
	\hspace{2cm}
	
	\pchspace
	\procedure{$\Exp{\UE,\adv}{\EUmod{1}}$}{%
		\pcln \textbf{do} ~\mathbf{Setup}(1^{n})\\
		\pcln \tx{b}'\gets \adv^{\tx{oracles}(\tx{b}=1)}(1^{n})\\
		\pcln\pcif \Set{K}^{*}\cap \Set{C}^{*}\neq \emptyset~\textbf{then}\\
		\pcln~~~~~\tx{twf}\gets 1\\
		\pcln\pcif \tx{twf}= 1~\textbf{then}\\
		\pcln ~~~~~\tx{b}'\sample \bin\\
		\pcln   \pcreturn \tx{b}' }		
\end{pchstack}

\vspace{0.5cm}

Although our $\UE$ scheme is a general randomized construction, it can be readily reformulated as a deterministic one. Nevertheless, in this work, we focus exclusively on the randomized setting and accordingly tailor the security model to this case. The security definition follows the standard semantic security approach, in which the adversary must distinguish between two experiments, namely $\Exp{\UE,\adv}{\EUmod{0}}$ and $\Exp{\UE,\adv}{\EUmod{1}}$. We emphasize that the proposed scheme achieves indistinguishability under chosen-plaintext attacks ($\tx{CPA}$).

%\noindent 
A $\UE$ scheme is said to be $\EUS$-secure if, for every PPT adversary $\adv$, the following advantage is negligible as a function of the security parameter $n$:
\begin{align*}
	\Adv_{\adv}^{\EUS}(n) := \abs{\Pr\left[ \Exp{\UE,\adv}{\EUmod{0}} = 1 \right] - \Pr\left[ \Exp{\UE,\adv}{\EUmod{1}} = 1 \right]}.
\end{align*}

%\vspace{0.3cm}
%\noindent
%The construction of our $\UE$ scheme is based on $\tx{FrodoPKE}$ learning with errors key encapsulation \cite{Frodo}.

\section{Frodo based UE scheme}
\subsection{Frodo KEM}\label{sec::Frodo}
Since the construction of our $\UE$ scheme is based on $\tx{FrodoPKE}$, we begin by presenting the framework for the Frodo learning with errors key encapsulation mechanism~\cite{Frodo, Frodo2}. To that end, let $D \in \ZZ_{>0}$ and define $q = 2^D$, with $B \in \ZZ_{[0, D]}$, and $n, \bar{m}, \bar{n} \in \ZZ_{>0}$, where $n \equiv 0 \pmod{8}$. Additionally, let $\Set{M} = \{0,1\}^{\ell}$ denote the message space (see below).

The error distribution $\chi$ used in $\tx{FrodoPKE}$ is a discrete, symmetric distribution over $\ZZ$, centered at zero with small support. It approximates a rounded continuous Gaussian distribution. The support of $\chi$ is defined as $S_{\chi} = \{-s, -s+1, \ldots, -1, 0, 1, \ldots, s-1, s\}$ for some $s \in \ZZ_{>0}$. The probabilities satisfy $\chi(z) = \chi(-z)$ for all $z \in S_{\chi}$ and are determined by a discrete probability density function represented as a table $T_{\chi} = (T_{\chi}(0), T_{\chi}(1), \ldots, T_{\chi}(s))$ of $s+1$ positive integers, which correspond to cumulative distribution function. Sampling from $\chi$ is performed using inversion sampling techniques; see~\cite{Frodo, Frodo2} for full details.
The distribution $\chi$ plays a central role in the algorithm $\tx{Frodo.SampleMatrix}$ (see~\cite{Frodo, Frodo2}), which is used to sample error matrices. For notational simplicity, the output of this algorithm will be denoted as $\vr{E} \sample \chi(\ZZ_q^{\dim})$ throughout the paper.

The public-key encryption scheme $\tx{FrodoPKE}$ serves as a foundational component of our construction and, as mentioned above, is based on the computational hardness of the $\LWE$ problem. This assumption provides strong security guarantees against both classical and quantum adversaries. $\tx{FrodoPKE}$ follows the standard architecture of lattice-based encryption schemes and consists of four PPT algorithms. Formally, it is defined as the following tuple: $\tx{FrodoPKE}$ is defined as a tuple of algorithms: $(\tx{FrodoPKE.Setup},\ \tx{FrodoPKE.KeyGen},$ $\tx{FrodoPKE.Enc},\ \tx{FrodoPKE.Dec})$,
each of which is described in detail below.

\vspace{0.5cm}
 \begin{pchstack}[center]
	
	\begin{pcvstack}
		
		\procedure{$\tx{FrodoPKE.Setup}(1^{n})$}{
			\pcln \vr{A}\sample \ZZ_{q}^{n\times n}\\
			\pcln \params\gets (\vr{A}, 1^{n},D, B, n,\bar{m}, \bar{n})\\
			\pcln \pcreturn \params}
		
		\pcvspace
		
		\procedure{$\tx{FrodoPKE.KeyGen}(\params)$}{
			\pcln \vr{S}, \vr{E} \sample \chi(\ZZ_{q}^{n\times \bar{n}})\\
			\pcln \vr{B} \gets \vr{A}\cdot \vr{S} + \vr{E} \in \ZZ_{q}^{n\times \bar{n}}\\
			\pcln \pcreturn \pk\gets \vr{B}, ~\sk\gets \vr{S}}

	\end{pcvstack}
	
	\pchspace
	
	\begin{pcvstack}
		~~~~~
     \end{pcvstack}	
		
	\pchspace	
	
	\begin{pcvstack}
	
	\procedure{$\tx{FrodoPKE.Enc}(\pk, \m\in \Set{M})$}{
		\pcln \vr{S}', \vr{E}' \sample \chi(\ZZ_{q}^{\bar{m}\times {n}})\\
		\pcln \vr{E}'' \sample \chi(\ZZ_{q}^{\bar{m}\times \bar{n}})\\
		\pcln \vr{B}' \gets \vr{S}' \cdot\vr{A} + \vr{E}' \in \ZZ_{q}^{\bar{m}\times {n}}\\
		\pcln \vr{V} \gets \vr{S}' \cdot\vr{B} + \vr{E}'' \in \ZZ_{q}^{\bar{m}\times \bar{n}}\\
		\pcln  \vr{C}_{1} \gets \vr{B}', ~\vr{C}_{2} \gets \vr{V} + \tx{encode}(\m)\\
		\pcln \pcreturn \ct\gets(\vr{C}_{1}, \vr{C}_{2})}
	
	\pcvspace  
	
	\procedure{$\tx{FrodoPKE.Dec}(\ct, \sk)$}{
		\pcln \vr{M} \gets \vr{C}_{2} - \vr{C}_{1}\cdot \vr{S}  \in \ZZ_{q}^{\bar{m}\times \bar{n}}\\
		\pcln \m'\gets \tx{decode}(\vr{M})\\
		\pcln\pcreturn \m'
	}
\end{pcvstack}	
\end{pchstack}	

 \vspace{0.5cm}
Although the functions $\tx{encode}$ and $\tx{decode}$ are formally defined in the Frodo specifications~\cite{Frodo, Frodo2}, we briefly outline the intuition behind their construction. The $\tx{encode}$ function maps bit strings of length $\ell = B \cdot \bar{m} \cdot \bar{n}$ to $\bar{m} \times \bar{n}$ matrices over $\ZZ_q$. To achieve this, each consecutive $B$-bit substring is interpreted as an integer $k \in [0, 2^B)$ and then mapped to an element of $\ZZ_q$ via the transformation $\tx{ec}(k) = k \cdot q / 2^B$. The corresponding $\tx{decode}$ function inverts this process using the operation $\tx{dc}(c) = \lfloor c \cdot 2^B / q \rceil \pmod{2^B}$.

The scheme is correct as long as the error remains sufficiently small.
\begin{lemma}(\cite{Frodo})\label{lem::01}
Let $q = 2^{D}$ and $B \leq D$. Then, for any integer $0 \leq k < 2^{B}$ and any error term $e$ satisfying $-q/2^{B+1} \leq e < q/2^{B+1}$, it holds that $\tx{dc}(\tx{ec}(k) + e) = k$.
\end{lemma}	

This result ensures that the decoding procedure remains robust to small additive errors introduced during encryption, as long as the noise stays within a carefully bounded interval. Intuitively, the rounding operation at the heart of $\tx{decode}$ acts as a form of noise tolerance, enabling perfect recovery of the original message bit string despite the presence of moderate $\LWE$ noise.

%The scheme is correct as long as the error remains sufficiently small.
%
%\begin{lemma}[\cite{Frodo}]\label{lem::01}
%	Let $q = 2^{D}$ and $B \leq D$. Then, for any integer $0 \leq k < 2^{B}$ and any error term $e$ satisfying $-q/2^{B+1} \leq e < q/2^{B+1}$, it holds that $\tx{dc}(\tx{ec}(k) + e) = k$.
%\end{lemma}

\section{Construction of  EU scheme based on Frodo}
To update a ciphertext $\ct_{\epoch}$, we homomorphically decrypt it using the update token $\tok_{\epoch+1}$ and re-encrypt the resulting message under the next epoch’s public key, yielding $\ct_{\epoch+1} \gets \tx{FrodoPKE.Enc}$ $(\pk_{\epoch+1}, \m)$. The update token $\tok_{\epoch+1}$ can be naturally viewed as a homomorphic encryption of the secret key $\sk_{\epoch}$ under the public key $\pk_{\epoch+1}$.
This transformation relies on two supporting functions: \emph{bit ordering} ($\tx{Ord}$) and \emph{Tensor-$D$} ($\bigotimes_{D}$), both of which are formally defined below. These functions are designed to enable the structured manipulation of ciphertexts necessary for homomorphic key updates within the lattice-based framework.
The overall approach is motivated by the key-switching technique originally introduced by Brakerski et al.~\cite{brakerski2011fully, brakerski2014leveled, brakerski2014efficient}, which enables ciphertexts to be transformed from one encryption key to another. It also draws on the noise smudging lemma~\cite{asharov2012multiparty, nishimaki2022direction}, which is essential for ensuring that the update process remains secure and error-resilient.

\begin{description}
	
	\item[{Bit-ordering}] {($\tx{Ord}$)}. This function takes as input a matrix $\vr{X}\in \ZZ_{q}^{\bar{m}\times \bar{n}}$ and produces an output matrix  $\vr{Y}\in \bin^{\bar{m}\times \bar{n}D}$. Let $\vr{X}[i,\cdot]=[{x}_{i1}, {x}_{i2}, \ldots , {x}_{i\bar{n}}]$ denote the $i$-th row of $\vr{X}$. Each entry ${x}_{i,j}\in\ZZ_{q}$ is represented using $D$ bits, meaning it can be written as ${x}_{i,j}=\sum_{k=1}^{D}2^{k-1}y_{i,j,k}$. The operator $\tx{Ord}$ transforms the row $\vr{X}[i,\cdot]$ into the $i$-th row of $\vr{Y}$, given by $\vr{Y}[i,\cdot]=[\vr{y}_{i1}, \vr{y}_{i2}, \ldots , \vr{y}_{iD}]$, where each $\vr{y}_{i,k}=[y_{i,j,k}]_{j}\in\bin^{\bar{n}}$.
	
	\item[Tensor-$D$] ($\bigotimes_{D}$  ). This function transforms a matrix from $\ZZ_{q}^{n\times \bar{n}}$ to $\ZZ_{q}^{nD\times \bar{n}}$. Let $\vr{X} =[\vr{x}_{1}, \vr{x}_{2}, \ldots , \vr{x}_{\bar{n}}]\in \ZZ_{q}^{n\times \bar{n}}$, where $\vr{x}_{j}$ represents the $j$-th column of $\vr{X}$. The transformation $\bigotimes_{D}(\vr{X}) = \vr{Y}=[\vr{y}_{1}, \vr{y}_{2}, \ldots , \vr{y}_{\bar{n}}]\in \ZZ_{q}^{nD\times \bar{n}}$, where each column $\vr{y}_{j}$ is defined as $\vr{y}_{j}=[1, 2, \ldots , 2^{D-1}]^{T}\otimes \vr{x}_{j}$.
	
	\item[Formal product] Consider matrices $\mathbf{A} \in \mathbb{Z}_q^{l \times r}$, $\mathbf{B} \in \mathbb{Z}_q^{r \times s_1}$, and $\mathbf{C} \in \mathbb{Z}_q^{r \times s_2}$. The formal product $\mathbf{A} \cdot (\mathbf{B}, \mathbf{C})$ represents the multiplication of $\mathbf{A}$ with the concatenation of $\mathbf{B}$ and $\mathbf{C}$, denoted as $\mathbf{A} \cdot [\mathbf{B} \mid \mathbf{C}]$. This operation yields $[\mathbf{A} \cdot \mathbf{B} \mid \mathbf{A} \cdot \mathbf{C}]$. By partitioning the resulting matrix along the concatenation, we establish the formal equivalence $(\mathbf{A} \cdot \mathbf{B}, \mathbf{A} \cdot \mathbf{C}) = \mathbf{A} \cdot (\mathbf{B}, \mathbf{C})$.
\end{description}

\begin{remark}
Note that if $\vr{C}$ and $\vr{S}$ belong to the domains of Bit-ordering and Tensor-$D$, respectively, then the following equality holds
$\tx{Ord}(\vr{C})\cdot \bigotimes_{D} (\vr{S}) = \vr{C}\cdot \vr{S}$.
\end{remark}

With a comprehensive understanding of the {bit-ordering}, {tensor-$D$} and {formal product} functions established, we can now delve into the specifics of our $\UE$ scheme. These foundational components are integral to the scheme's architecture, enabling efficient key updates and ensuring the continuous security of encrypted data. The following exposition will detail the design principles, operational procedures, and the interplay of these core functions that facilitate seamless and secure encryption transitions.

\vspace{0.5cm}
 \begin{pchstack}[center]	
	\begin{pcvstack}
		\procedure{$\tx{UE.Setup}(1^{n})$}{
			\pcln \params\sample \tx{FrodoPKE.Setup}(1^{n})\\
			\pcln \params=(\vr{A}, 1^{n},D, B, n,\bar{m}, \bar{n})\\
			\pcln\pcreturn \params}
		
		\pcvspace
		
		\procedure{$\uekg(\params)$}{
			\pcln (\vr{B}_{\epoch}, \vr{S}_{\epoch})  \sample \tx{FrodoPKE.KeyGen}(\params)\\
			\pcln \sk_{\epoch}\gets \vr{S}_{\epoch},~\pk_{\epoch} \gets \vr{B}_{\epoch}\\
			\pcln\pcreturn \key_{\epoch} \gets (\sk_{\epoch}, \pk_{\epoch})}
		
	\end{pcvstack}
	
	\pchspace
	\hspace{0.7cm}
	
	\begin{pcvstack}
		\procedure{$\ueenc(\key_{\epoch}, \m\in\Set{M})$}{
			\pcln   \key_{\epoch} = (\vr{S}_{\epoch}, \vr{B}_{\epoch}) \\
			\pcln (\vr{C}_{1}, \vr{C}_{2}) \sample \tx{FrodoPKE.Enc}(\vr{B}_{\epoch}, \m)\\
			\pcln  \ct\gets(\vr{C}_{1}, \vr{C}_{2}) \in  \ZZ_{q}^{\bar{m}\times {n}}  \times \ZZ_{q}^{\bar{m}\times \bar{n}}\\
			\pcln\pcreturn  \ct}
		
		\pcvspace
		
		\procedure{$\uedec(\key_{\epoch}, \ct)$}{
			\pcln   \key_{\epoch} = (\vr{S}_{\epoch}, \vr{B}_{\epoch}), ~\ct = (\vr{C}_{1}, \vr{C}_{2})  \\
			\pcln \m' \gets\tx{FrodoPKE.Dec}(\ct, \vr{S}_{\epoch})\\
			\pcln\pcreturn  \m'}
	\end{pcvstack}
\end{pchstack}	

\begin{pcvstack}[center]	
	\procedure{$\uetg(\key_{\epoch}, \key_{\epoch+1})$}{
		\pcln   \key_{\epoch} = (\vr{S}_{\epoch}, \vr{B}_{\epoch}), ~  \key_{\epoch+1} = (\vr{S}_{\epoch+1}, \vr{B}_{\epoch+1}) \\
		\pcln  \vr{S}_{(1)}', \vr{E}_{(1)}' \sample \chi(\ZZ_{q}^{nD\times {n}}), ~ \vr{E}_{(1)}'' \sample \chi(\ZZ_{q}^{nD\times \bar{n}}) \\
		\pcln \vr{S}_{(2)}', \vr{E}_{(2)}' \sample \chi(\ZZ_{q}^{n\times {n}}), ~ \vr{E}_{(2)}'' \sample \chi(\ZZ_{q}^{n\times \bar{n}}) \\
		\pcln \tok_{\epoch+1}^{(1)} \gets \left( \vr{S}_{(1)}'\cdot\vr{A}  +  \vr{E}_{(1)}',  \vr{S}_{(1)}'\cdot\vr{B}_{\epoch+1}+\vr{E}_{(1)}''  - {{\scriptsize\bigotimes}_{D}}(\vr{S}_{\epoch}) \right) \\
		\pcln \tok_{\epoch+1}^{(2)} \gets \left( \vr{S}_{(2)}'\cdot\vr{A}  +  \vr{E}_{(2)}',  \vr{S}_{(2)}'\cdot\vr{B}_{\epoch+1}+\vr{E}_{(2)}''\right)    \\
		\pcln\pcreturn  \tok_{\epoch+1}\gets (  \tok_{\epoch+1}^{(1)} ,  \tok_{\epoch+1}^{(2)} )}
\end{pcvstack}		

\vspace{0.5cm}
%%%%

\begin{pcvstack}[center]	
	%\begin{pcvstack}[center]		
	\procedure{$\ueupd( \tok_{\epoch+1}, \ct_{\epoch})$}{
		\pcln  \tok_{\epoch+1} = (  \tok_{\epoch+1}^{(1)} ,  \tok_{\epoch+1}^{(2)} ), ~ \ct_{e}=(\vr{C}_{\epoch,1}, \vr{C}_{\epoch,2}) ~~~~~~~~~~~~~~~~~~~~~~~~~~~~~~\\
		\pcln (\vr{C}_{1}^{(1)},  \vr{C}_{2}^{(1)}) \gets \tx{Ord}(\vr{C}_{\epoch,1})\cdot \tok_{\epoch+1}^{(1)}\\
		\pcln \vr{R} \sample \chi(\ZZ_{q}^{\bar{m}\times {n}})\\
		\pcln (\vr{C}_{1}^{(2)},  \vr{C}_{2}^{(2)}) \gets \vr{R}\cdot  \tok_{\epoch+1}^{(2)}\\
		\pcln \ct_{\epoch+1}\gets \left(  \vr{C}_{1}^{(1)} + \vr{C}_{1}^{(2)}, \vr{C}_{\epoch,2} + \vr{C}_{2}^{(1)} + \vr{C}_{2}^{(2)}\right)\\
		\pcln\pcreturn  \ct_{\epoch+1} }
	
\end{pcvstack}		

\vspace{0.5cm}
\begin{remark}\label{rem::01}
	Note that to generate the token $\tok_{\epoch+1}$, $\uetg$ only needs $\sk_{\epoch}$ and $\pk_{\epoch+1}$, rather than the complete keys $\key_{\epoch}$ and $\key_{\epoch+1}$. This aligns with the idea introduced earlier.
\end{remark}

The following theorem outlines the conditions under which the scheme operates correctly.
%The scheme is correct. Indeed, 
%\begin{theorem}
%	Let $\chi$ be an error distribution defined in Sec. \ref{sec::Frodo} such that $S_{\chi}=\{ -s, \ldots, 0,  \ldots ,s \}$ is its support, and
%	let $T$ be the maximum number of the epochs. If $2(n^{2}Ds^{3}+n^{2}s^{3})+nDs + ns^{2}<q/(T\cdot 2^{B+1})$, then the scheme is correct.
%\end{theorem}	
\begin{theorem}\label{th::06::03}
	Suppose $\chi$ is the error distribution defined in Sec.~\ref{sec::Frodo}, with support $S_{\chi} = \{ -s, \ldots, 0, \ldots, s \}$. Assume $T$ is the maximum number of epochs. If
	\[
	2(n^{2}Ds^{3} + n^{2}s^{3}) + nDs + ns^{2} < \frac{q}{T \cdot 2^{B+1}},
	\]
	then the scheme is correct.
\end{theorem}

\begin{proof}
Let $\ct_{\epoch} =  (\vr{C}_{\epoch,1}, \vr{C}_{\epoch, 2})$ and $\ct_{\epoch+1} =  (\vr{C}_{\epoch+1,1}, \vr{C}_{\epoch+1, 2})$, where the components are defined as in $\ueupd$. It is straightforward to verify that 
 \begin{align*}
	\vr{C}_{\epoch+1,2} - \vr{C}_{\epoch +1, 1}\cdot \vr{S}_{\epoch +1} = \vr{C}_{\epoch,2} - \vr{C}_{\epoch , 1}\cdot \vr{S}_{\epoch } + \tx{Error}
\end{align*}
where the error term is given by
\begin{align}\label{pr01::01}
	\tx{Error}= \left(\tx{Ord}(\vr{C}_{\epoch , 1})\cdot\vr{S}_{(1)}' + \vr{R}\cdot \vr{S}_{(2)}'\right)\vr{E}- \left(\tx{Ord}(\vr{C}_{\epoch , 1})\cdot\vr{E}_{(1)}' + \vr{R}\cdot \vr{E}_{(2)}'\right)\vr{S}_{\epoch+1} \\
	+ \tx{Ord}(\vr{C}_{\epoch , 1})\cdot\vr{E}_{(1)}'' + \vr{R}\cdot \vr{E}_{(2)}'' \nonumber
\end{align}
Here, the primed variables involving $\mathbf{E}$ (i.e., $\mathbf{E}'_{(1)}, \mathbf{E}'_{(2)}, \mathbf{E}''_{(1)}, \mathbf{E}''_{(2)}$) denote independent noise samples generated within $\mathsf{UE.TG}$, following the structure of $\LWE$-based encryption. The matrix $\mathbf{E}$ denotes the noise introduced when generating (or, in this case, updating) $\mathsf{pk}_{\epoch+1} = \mathbf{B}_{\epoch+1}$ from $\mathsf{sk}_{\epoch+1} = \mathbf{A} \cdot \mathbf{S}_{\epoch+1}$, in the same form as described in $\mathsf{FrodoPKE.KeyGen}$.

\noindent Given that $\tx{Ord}(\vr{C}_{\epoch , 1})\in\{0,1\}^{\bar{m}\times nD}$, $\vr{S}_{(1)}' \in S_{\chi}^{nD\times n}$, and $\vr{E} \in S_{\chi}^{n\times n}$, we can derive the following bound
\begin{align}\label{pr01::02}
	\left\|  \left(\tx{Ord}(\vr{C}_{\epoch , 1})\cdot\vr{S}_{(1)}' + \vr{R}\cdot \vr{S}_{(2)}'\right)\vr{E} \right\|_{\max}\leq n^{2}Ds^{2}+n^{2}s^{3}.
\end{align}
Similarly, we have
\begin{align}
	\left\|  \left(\tx{Ord}(\vr{C}_{\epoch , 1})\cdot\vr{E}_{(1)}' + \vr{R}\cdot \vr{E}_{(2)}'\right)\vr{S}_{\epoch+1}    \right\|_{\max}\leq n^{2}Ds^{2}+n^{2}s^{3}, \label{pr01::03}\\
		\left\|  \tx{Ord}(\vr{C}_{\epoch , 1})\cdot\vr{E}_{(1)}'' \right\|_{\max}\leq nDs, ~~~ \left\|  \vr{R}\cdot \vr{E}_{(2)}'' \right\|_{\max}\leq ns^{2}. \label{pr01::04}
\end{align}
By applying the triangle inequality to (\ref{pr01::01}) together with the bounds in (\ref{pr01::02})–(\ref{pr01::04}), we obtain:
\begin{align*}
	\norm{\tx{Error}}_{\infty}\leq 2(n^{2}Ds^{3}+n^{2}s^{3})+nDs + ns^{2}<q/(T\cdot 2^{B+1}).
\end{align*}
Therefore, each ciphertext update introduces noise of magnitude at most $q / (T \cdot 2^{B+1})$. It follows that after $T$ updates, the total accumulated noise remains below $q / 2^{B+1}$. Hence, by Lemma~\ref{lem::01}, the proof is complete.
\end{proof}	

\subsection{Security proof}
%In this section we justify that the scheme is $\EUS$-secure in the backward-leak uni-directional setting. To this end, we follow the firewall technique \cite{jiang2020,nishimaki2022direction}.
%
%Let us start with the explanation that the scheme meets, in particular, uni-directional ciphertext updates from its security.
%\begin{theorem}
%	The advantage of any PPT adversary $\adv$ in converting a ciphertext under a public key $\pk_{\epoch+1}$ into a ciphertext under $\pk_{\epoch}$ by using $(\sk_{\epoch},\pk_{\epoch}), \pk_{\epoch+1}$ and $\tok_{\epoch+1}$ is negligible.
%\end{theorem}	
In this section, we prove that the scheme satisfies $\EUS$-security in the backward-leak, uni-directional setting. This security notion ensures that even if secret keys from earlier epochs are leaked, an adversary cannot compromise ciphertexts encrypted under newer public keys. Our proof adopts the \emph{firewall technique}, a methodology introduced by Jiang and Nishimaki~\cite{jiang2020,nishimaki2022direction}, which provides a systematic framework for analyzing security in key-evolving encryption schemes.

As a first step, we show that the scheme inherently supports uni-directional ciphertext updates, an essential property that follows directly from its security guarantees.

\begin{theorem}
	For any PPT adversary $\adv$, the advantage of converting a ciphertext under public key $\pk_{\epoch+1}$ into a valid ciphertext under $\pk_{\epoch}$, given access to $(\sk_{\epoch}, \pk_{\epoch})$, $\pk_{\epoch+1}$, and $\tok_{\epoch+1}$, is negligible.
\end{theorem}

\begin{proof}
Suppose, for contradiction, that the adversary $\adv$ is capable of transforming a ciphertext under $\pk_{\epoch+1}$ into a valid ciphertext under $\pk_{\epoch}$. Given that $\tx{FrodoPKE}$ is $\indcpa$-secure, we construct a PPT adversary $\adversary{B}$ that breaks the security of ciphertexts under $\pk_{\epoch+1}$.  

The adversary $\adversary{B}$ is given the public parameters $\params$ and the public key $\pk_{\epoch+1}$. It then generates a key pair $(\sk_{\epoch}, \pk_{\epoch})\gets \uekg(\params)$ and computes the update token $\tok_{\epoch+1}\gets\uetg(\sk_{\epoch},\pk_{\epoch+1})$ (see Remark \ref{rem::01}).
Next, $\adversary{B}$ selects two equal-length messages $\m_{0}, \m_{1}\in\Set{M}$ and submits them to its $\indcpa$ challenger.
The challenger randomly selects a bit $b \sample \bin$, encrypts $\m_b$ as $\ct^* \gets \tx{FrodoPKE.Enc}(\pk_{\epoch+1}, \m_b)$, and returns the challenge ciphertext $\ct^*$ to $\adversary{B}$. At this point, $\adversary{B}$ runs $\adv$ on input $((\sk_{\epoch},\pk_{\epoch}),\tok_{\epoch+1}, \ct^{*})$. 
The adversary $\adv$ outputs a ciphertext $\ct'$ under $\pk_{\epoch}$. Then, $\adversary{B}$ decrypts this ciphertext by computing $\m'\gets \tx{FrodoPKE.Dec}(\ct', \sk_{\epoch})$ and outputs a guess $b'$ such that $\m' = \m_{b'}$. Since $\adv$ successfully converts the ciphertext with overwhelming probability, it follows that $b' = b$ with non-negligible advantage, contradicting the assumed $\indcpa$-security of $\tx{FrodoPKE}$. This completes the proof.
\end{proof}	

In the next step, we reformulate the security game and propose new hybrid algorithms that simulate the scheme with negligible error, while remaining easier to analyze. To proceed, we revisit the update procedure. By the scheme’s definition, we have:

\begin{align*}
	 \vr{C}_{1}^{(1)} &= \tx{Ord}(\vr{C}_{\epoch , 1})\cdot\vr{S}_{(1)}'  \cdot \vr{A} + \tx{Ord}(\vr{C}_{\epoch , 1})\cdot \vr{E}_{(1)}' ,     \\
	 \vr{C}_{2}^{(1)} &=   \tx{Ord}(\vr{C}_{\epoch , 1})\cdot \vr{S}_{(1)}' \cdot \vr{B}_{\epoch+1} + \tx{Ord}(\vr{C}_{\epoch , 1})\cdot \vr{E}_{(1)}''  -   \underbrace{\tx{Ord}(\vr{C}_{\epoch , 1})\cdot \bigotimes_{D} (\vr{S}_{\epoch}) }_{\vr{C}_{\epoch , 1}\cdot \vr{S}_{\epoch}}, \\
	 	\vr{C}_{1}^{(2)} &= \vr{R}\cdot \vr{S}_{(2)}' \cdot \vr{A} + \vr{R}\cdot \vr{E}_{(2)}'  ,\\
	 \vr{C}_{2}^{(2)} &= \vr{R}\cdot \vr{S}_{(2)}' \cdot \vr{B}_{\epoch+1} + \vr{R}\cdot \vr{E}_{(2)}''.
\end{align*}
A straightforward calculation yields
\begin{align*}
 \vr{C}_{1}^{(1)} + \vr{C}_{1}^{(2)}  = \left(  \underbrace{\tx{Ord}(\vr{C}_{\epoch , 1})\cdot \vr{S}_{(1)}'  +  \vr{R}\cdot \vr{S}_{(2)}' }_{\vr{S}^{\dagger}}  \right) \cdot \vr{A}  + \underbrace{\tx{Ord}(\vr{C}_{\epoch , 1})\cdot \vr{E}_{(1)}' + \vr{R}\cdot \vr{E}_{(2)}' }_{\vr{E}^{\dagger}} ,
\end{align*}
and
\begin{align*}
 \vr{C}_{2}^{(1)}  +  \vr{C}_{2}^{(2)}  + \vr{C}_{\epoch,2}  = & \left(  \underbrace{\tx{Ord}(\vr{C}_{\epoch , 1})\cdot \vr{S}_{(1)}'  +  \vr{R}\cdot \vr{S}_{(2)}' }_{\vr{S}^{\dagger}}  \right) \cdot \vr{B}_{\epoch+1} \\
 & + \underbrace{\tx{Ord}(\vr{C}_{\epoch , 1})\cdot \vr{E}_{(1)}'' + \vr{R}\cdot \vr{E}_{(2)}'' +\vr{E}_{\epoch}''}_{\vr{E}^{\dagger\dagger}} \\
 &- \underbrace{\vr{E}'\cdot \vr{S}_{\epoch} + \vr{S}' \cdot \vr{E}}_{\text{''garbage''}}~ +  ~\tx{encode}(\m)
\end{align*}
The primed variants of $\mathbf{E}$ represent independent noise samples generated within $\mathsf{UE.TG}$, while $\mathbf{E}$ refers to the noise used in the public key update. See the explanation in the proof of Theorem~\ref{th::06::03} for further context.

\vspace{0.3cm}
\noindent This leads us to the following conclusion.
\begin{concl}\label{conc::01}
	By Lemma \ref{lem::smudging},  a ciphertext $\ct_{\epoch+1}$ is statistically indistinguishable from $(\vr{S}^{\dagger}\cdot\vr{A} + \vr{E}^{\dagger},  \vr{S}^{\dagger}\cdot\vr{B}_{\epoch+1} + \vr{E}^{\dagger\dagger} +  \tx{encode}(\m))$. This implies that we can simulate an updated ciphertext using the original ciphertext, its associated plaintext and randomness, the new epoch's public key, and fresh randomness for generating the token $\tok_{\epoch+1}$.
\end{concl}

\vspace{0.5cm}
%\begin{pcvstack}[center]	

\procedure{$\tx{Hyb.UE.Upd}(\ct_{\epoch}, \vr{B}_{\epoch+1},\m, \vr{E}_{\epoch}'', (\vr{S}_{(1)}', \vr{S}_{(2)}', \vr{E}_{(1)}'', \vr{E}_{(2)}''))$}{
	\pcln \text{Parse} ~~ \ct_{\epoch} = (\vr{C}_{\epoch, 1}, \vr{C}_{\epoch,2})\\
	\pcln \text{Choose} ~~ \vr{R} \sample \chi(\ZZ_{q}^{\bar{m}\times {n}})\\
	\pcln \text{Set} ~~ \vr{S}^{\dagger} = \tx{Ord}(\vr{C}_{\epoch , 1})\cdot \vr{S}_{(1)}'  +  \vr{R}\cdot \vr{S}_{(2)}' \\
	\pcln \text{Set} ~~ \ct_{\epoch+1} =  \left(\vr{S}^{\dagger}\cdot\vr{A} + \vr{E}^{\dagger},  \vr{S}^{\dagger}\cdot\vr{B}_{\epoch+1} + \vr{E}^{\dagger\dagger} +  \tx{encode}(\m)\right) \\
	\pcln \text{Output} ~~ \left( \ct_{\epoch+1} ,  \vr{E}_{\epoch}''  \right).
}

\vspace{0.3cm}

\procedure{$\tx{Sim.UE.KG}(\params)$}{
	\pcln \text{Choose} ~~ \vr{B}_{e}^{+} \sample  \ZZ_{q}^{n\times \bar{n}} \\
	\pcln \text{Output} ~~ \pk_{\epoch} = \vr{B}_{e}^{+}.}

\vspace{0.3cm}

\procedure{$\tx{Sim.UE.TG}(\params)$}{
	\pcln \text{Choose} ~~\tok_{\epoch+1}^{(1)+} \sample  \ZZ_{q}^{nD\times {n}} \times \ZZ_{q}^{nD\times \bar{n}},~  \text{and} ~ \tok_{\epoch+1}^{(2)+}  \sample    \ZZ_{q}^{n\times {n}} \times \ZZ_{q}^{D\times \bar{n}}  \\
	\pcln \text{Output} ~~ \tok_{\epoch+1}^{+} = \left( \tok_{\epoch+1}^{(1)+},   \tok_{\epoch+1}^{(2)+}  \right).}

\vspace{0.3cm}

\procedure{$\tx{Sim.UE.Upd}(\params)$}{
	\pcln \text{Choose} ~~  (\vr{C}_{1}, \vr{C}_{2}) \sample  \ZZ_{q}^{\bar{m}\times {n}}  \times \ZZ_{q}^{\bar{m}\times \bar{n}} \\
	\pcln \text{Output} ~~ \ct_{\epoch+1} = (\vr{C}_{1}, \vr{C}_{2}).}

\vspace{0.3cm}

\procedure{$\tx{Sim.UE.Enc}(\params)$}{
	\pcln \text{Choose} ~~  (\vr{C}_{1}, \vr{C}_{2}) \sample  \ZZ_{q}^{\bar{m}\times {n}}  \times \ZZ_{q}^{\bar{m}\times \bar{n}} \\
	\pcln \text{Output} ~~ \ct_{\epoch} = (\vr{C}_{1}, \vr{C}_{2}).}

%\end{pcvstack}	(\vr{C}_{1}, \vr{C}_{2}) \in  \ZZ_{q}^{\bar{m}\times {n}}  \times \ZZ_{q}^{\bar{m}\times \bar{n}}

\vspace{1cm}

\noindent It follows directly from Conclusion~\ref{conc::01} that the following lemma holds.
\begin{lemma}
	The ciphertext update  $\ueupd( \tok_{\epoch+1}, \ct_{\epoch})$ is statistically indistinguishable form \\$\tx{Hyb.UE.Upd}(\ct_{\epoch}, \vr{B}_{\epoch+1},\m, \vr{E}_{\epoch}'', (\vr{S}_{(1)}', \vr{S}_{(2)}', \vr{E}_{(1)}'', \vr{E}_{(2)}''))$.	
\end{lemma}

\noindent From the lemma above, we derive the following conclusion.
\begin{concl}\label{conc::02}
	The oracle $\Upd(\ct_{\epoch})$ can be simulated by $\tx{Hyb.UE.Upd}(\ct_{\epoch}, \vr{B}_{\epoch+1},\m, \vr{E}_{\epoch}'', (\vr{S}_{(1)}',$ $\vr{S}_{(2)}', \vr{E}_{(1)}'', \vr{E}_{(2)}''))$.	
\end{concl}

Let $T+1$ be a number of epochs (they are counted from zero). Define the following hybrid games: 
\begin{description}
	\item[$\tx{Hyb}_{i}(b):$] This is the same as $\Exp{\UE,\adv}{\EUmod{b}}(n)$ except the following difference: When the adversary sends a query $(\bar{\m}, \bar{\ct})$ to $\Chall$ or an empty query to $\UpdC$ at epoch $j$:
	\begin{itemize}
		\item for $j<i$ return an honestly generated challenge-equal ciphertext, i.e. 
		\begin{itemize}
			\item[{}] \textbf{if} $b=0$ \textbf{then} $\ueenc(\key_{\widetilde{\epoch}}, \bar{\m})$,
			\item[{}] \textbf{else} $\tx{UE.Upd}(\tok_{\widetilde{\epoch}}, \bar{\ct})$. 
		\end{itemize}
		\item $j\geq i$, return a random ciphertext.
	\end{itemize}
\end{description}
It is easy to see that $\tx{Hyb}_{T+1}(b)$ is the same as the original $\tx{rand\md ind\md eu\md cpa}$	game in the backward-leak uni-directional setting $\Exp{\UE,\adv}{\EUmod{b}}(n)$. Let $\mathcal{U} (n)$ be a random variable distributed uniformly in $[0,T]$, by the standard hybrid argument, we have
\begin{align*}
	\Adv_{\adv}&^{\EUS}(n) = \left| \Pr\left[ \tx{Hyb}_{T+1}(1)=1 \right]   -  \Pr\left[ \tx{Hyb}_{T+1}(0)=1 \right]   \right| \nonumber\\
	=&\left|  \sum_{i=0}^{T} \left\{   \left(  \Pr\left[   \tx{Hyb}_{i+1}(1)=1 \right]   -   \Pr\left[   \tx{Hyb}_{i}(1)=1 \right] \right) \right. \right. \nonumber\\
	& -   \left.\left. \left(  \Pr\left[   \tx{Hyb}_{i+1}(0)=1 \right]   -   \Pr\left[   \tx{Hyb}_{i}(0)=1 \right] \right)   \right\} \right|  
\end{align*}	
\begin{align}
	=&\left|  \sum_{i=0}^{T}  \left(  \Pr\left[   \tx{Hyb}_{\mathcal{U}(n)+1}(1)=1 \mid \mathcal{U}(n) = i \right]   -   \Pr\left[   \tx{Hyb}_{\mathcal{U}(n)}(1)=1 \mid \mathcal{U}(n) = i \right] \right) \right. \nonumber \\
	&\left. - \sum_{i=0}^{T}  \left(  \Pr\left[   \tx{Hyb}_{\mathcal{U}(n)+1}(0)=1 \mid \mathcal{U}(n) = i \right]   -   \Pr\left[   \tx{Hyb}_{\mathcal{U}(n)}(0)=1 \mid \mathcal{U}(n) = i \right] \right) \right| \label{eq::02} \\
	= & (T+1)\left|  \sum_{i=0}^{T}  \left(  \Pr\left[   \tx{Hyb}_{\mathcal{U}(n)+1}(1)=1 \land \mathcal{U}(n) = i \right]   -   \Pr\left[   \tx{Hyb}_{\mathcal{U}(n)}(1)=1 \land \mathcal{U}(n) = i \right] \right) \right. \nonumber\\ %\label{eq::03}
	&\left. - \sum_{i=0}^{T}  \left(  \Pr\left[   \tx{Hyb}_{\mathcal{U}(n)+1}(0)=1 \land \mathcal{U}(n) = i \right]   -   \Pr\left[   \tx{Hyb}_{\mathcal{U}(n)}(0)=1 \land \mathcal{U}(n) = i \right] \right) \right| \nonumber \\
	&\leq (T+1) \left|  \Pr\left[   \tx{Hyb}_{\mathcal{U}(n)+1}(1)=1\right] - \Pr\left[   \tx{Hyb}_{\mathcal{U}(n)}(1)=1  \right]  \right|  \nonumber \\
	&+  (T+1) \left|  \Pr\left[   \tx{Hyb}_{\mathcal{U}(n)+1}(0)=1\right] - \Pr\left[   \tx{Hyb}_{\mathcal{U}(n)}(0)=1  \right]  \right| \nonumber
\end{align}	
The condition $\tx{Hyb}_{0}(0)  =  \tx{Hyb}_{0}(1)$ trivially holds since all challenge-equal ciphertexts are random ciphertexts. We use this fact in (\ref{eq::02}). 

By the above computation, we have to prove that 
\begin{align*}
	\left|  \Pr\left[   \tx{Hyb}_{\mathcal{U}(n)+1}(b)=1\right] - \Pr\left[   \tx{Hyb}_{\mathcal{U}(n)}(b)=1  \right]  \right| \leq \negl, ~~~\text{where}~b\in\bin.
\end{align*}

In order to use the firewall technique \cite{jiang2020,nishimaki2022direction}, we define the next hybrid game:
\begin{description}
	\item[$\tx{Hyb}_{i}'(b):$] This is the same as $\tx{Hyb}_{i}(b)$ except that the game chooses $\tx{fwl},\tx{fwr}\gets [0,T]$. If the adversary corrupts $\key_{j}$ such that $j\in[\tx{fwl},\tx{fwr}]$ or $\tok_{\tx{fwr}+1}$, the game aborts.
\end{description}
The guess is correct with probability $(T+1)^{-2}$. Therefore, we have
\begin{align*}
	\left|  \Pr\left[   \tx{Hyb}_{\mathcal{U}(n)+1}(b)=1\right]\right. - & \left. \Pr\left[   \tx{Hyb}_{\mathcal{U}(n)}(b)=1  \right]  \right| \\
	&\leq (T+1)^{2}	\left|  \Pr\left[   \tx{Hyb}_{\mathcal{U}(n)+1}'(b)=1\right] - \Pr\left[   \tx{Hyb}_{\mathcal{U}(n)}'(b)=1  \right]  \right|.
\end{align*}
Note that to prove the security of our $\UE$ scheme it remains to show that the following condition holds.  
\begin{align}\label{eq::04}
	\left|  \Pr\left[   \tx{Hyb}_{\mathcal{U}(n)+1}'(b)=1\right] - \Pr\left[   \tx{Hyb}_{\mathcal{U}(n)}'(b)=1  \right]  \right|\leq \negl.
\end{align}

\noindent
Intuitively, this condition asserts that the adversary cannot distinguish between the final two hybrids in our sequence with more than negligible advantage. The lemma below formalizes this indistinguishability under the $\LWE$ assumption.
\begin{lemma}
	If the $\LWE$ assumption holds, then Equation~(\ref{eq::04}) also holds.
\end{lemma}
We omit the proof of this lemma, as it closely parallels the argument in Lemma~5.7 of Nishimaki’s work~\cite{nishimaki2022direction}, with only minor adaptations to suit our setting.

\eject

Our analysis leads to the following security guarantee for the proposed scheme.
\begin{theorem}\label{th::main::one}
	Under the $\LWE$ assumption, the presented $\tx{FrodoPKE}$-based $\UE$ scheme is $\tx{rand\md ind\md}$ $\tx{eu\md cpa}$	secure in the backward-leak uni-directional setting. That is, $\Adv_{\adv}^{\EUS}(n) \leq \negl$. 
\end{theorem}	

\noindent As stated, this result formally establishes the security of our scheme against both classical and quantum adversaries. While the underlying $\tx{FrodoPKE}$ construction is already known to be quantum-resistant due to its reliance on the $\LWE$ assumption, it was not a priori evident that the full updatable encryption scheme would inherit this level of security. The theorem above certifies that, despite the additional structure and operations introduced by the $\UE$ framework, the overall design remains robust in the post-quantum setting. This provides strong evidence that the scheme is suitable for deployment in environments requiring long-term confidentiality under realistic quantum threat models.

%\end{pcvstack}	(\vr{C}_{1}, \vr{C}_{2}) \in  \ZZ_{q}^{\bar{m}\times {n}}  \times \ZZ_{q}^{\bar{m}\times \bar{n}}

\section{Implementation and Performance Analysis}

This section presents a detailed performance analysis of the updatable encryption scheme, implemented using various configurations of $\tx{FrodoPKE}$. Specifically, we evaluate both AES-based and SHAKE-based instantiations of FrodoKEM at three different security levels defined by the {NIST Post-Quantum Cryptography (PQC) standardization project}, namely {security levels 1, 3, and 5}. These levels correspond to FrodoKEM-640, FrodoKEM-976, and FrodoKEM-1344, respectively.

The numerical suffixes in these names (640, 976, and 1344) refer to the matrix dimensions used in the scheme’s underlying lattice-based cryptographic operations. Each configuration is designed to meet a specific NIST PQC security level:
\begin{itemize}
	\item Level 1 (FrodoKEM-640) provides security approximately equivalent to AES-128,
	\item Level 3 (FrodoKEM-976) corresponds roughly to AES-192 security,
	\item Level 5 (FrodoKEM-1344) targets security comparable to AES-256.
\end{itemize}
Each security level supports two main variants, distinguished by the choice of pseudorandom number generator (PRNG): one using AES and the other using SHAKE (SHAKE128 or SHAKE256). While both variants offer the same level of security, they differ in performance characteristics. AES-based versions may benefit from hardware acceleration available on modern CPUs, whereas SHAKE-based versions are typically more portable and may be preferable in environments without AES-specific hardware support.

For each FrodoKEM variant, we measure and report the average execution time and standard deviation for key generation, encryption, decryption, token generation, and ciphertext update. These results provide a comprehensive view of the performance trade-offs associated with each parameter set and PRNG choice, helping to assess the practicality of deploying updatable encryption at different post-quantum security levels.

%\vspace{5mm}
\begin{table}[ht]
	\centering{
		\caption{\label{TestPhi} Execution times for the updatable encryption scheme based on $\tx{FrodoPKE}$ using AES.}
		\scalebox{1}{
			\hspace*{-2cm}
			\begin{tabular}{ |p{3.3cm}|p{1cm}|p{1cm}|p{1cm}|p{1cm}|p{1cm}|p{1cm}| }
				\hline
				\multicolumn{1}{|c|}{Algorithm variants} &\multicolumn{2}{|c|}{$\tx{FrodoPKE}$-640} & \multicolumn{2}{|c|}{$\tx{FrodoPKE}$-976} &  \multicolumn{2}{|c|}{$\tx{FrodoPKE}$-1344} \\
				\hline
				Algorithm part & Avg. time [s] & Std. & Avg. time [s] & Std. & Avg. time [s] & Std. \\
				\specialrule{.2em}{.1em}{.1em}
				UE.KG(params) & 0.0198 & 0.0006 & 0.0465 & 0.0030 & 0.0878 & 0.0035 \\
				\hline
				UE.Enc($\text{k}_e, \mathbf{m} \in \mathcal{M}$) & 0.0218 & 0.0011 & 0.0475 & 0.0010 & 0.0939 & 0.0010 \\
				\hline
				UE.Dec($\text{k}_e, \text{ct}_e$) & 0.0003 & 0.00001 & 0.0004 & 0.00001 & 0.0004 & 0.00001 \\
				\hline
				UE.TG($\text{k}_e, \text{k}_{e+1}$) & 6.3251 & 0.2018 & 18.0592 & 0.9919 & 64.1637 & 0.8593 \\
				\hline
				UE.Upd($\Delta_{e+1}, \text{ct}_e$) & 0.1224 & 0.0057 & 0.5130 & 0.0057 & 1.2693 & 0.0379 \\
				\hline
			\end{tabular}
		}
		\hspace*{-2cm}
	}
\end{table}
\begin{table}[ht]
	\centering{
		\caption{\label{TestPhi2} Execution times for the updatable encryption scheme based on $\tx{FrodoPKE}$ using SHAKE.}
		\scalebox{1}{
			\hspace*{-2cm}
			\begin{tabular}{ |p{3.3cm}|p{1cm}|p{1cm}|p{1cm}|p{1cm}|p{1cm}|p{1cm}| }
				\hline
				\multicolumn{1}{|c|}{Algorithm variants} &\multicolumn{2}{|c|}{$\tx{FrodoPKE}$-640} & \multicolumn{2}{|c|}{$\tx{FrodoPKE}$-976} &  \multicolumn{2}{|c|}{$\tx{FrodoPKE}$-1344} \\
				\hline
				Algorithm part & Avg. time [s] & Std. & Avg. time [s] & Std. & Avg. time [s] & Std. \\
				\specialrule{.2em}{.1em}{.1em}
				UE.KG(params) & 0.0198 & 0.0004 & 0.0471 & 0.0030 & 0.0894 & 0.0063 \\
				\hline
				UE.Enc($\text{k}_e, \mathbf{m} \in \mathcal{M}$) & 0.0218 & 0.0002 & 0.0484 & 0.0014 & 0.0945 & 0.0006 \\
				\hline
				UE.Dec($\text{k}_e, \text{ct}_e$) & 0.0002 & 0.00001 & 0.0004 & 0.00001 & 0.0005 & 0.00003 \\
				\hline
				UE.TG($\text{k}_e, \text{k}_{e+1}$) & 6.3118 & 0.0649 & 18.0746 & 0.4328 & 63.3840 & 0.8250 \\
				\hline
				UE.Upd($\Delta_{e+1}, \text{ct}_e$) & 0.1214 & 0.0012 & 0.5166 & 0.0153 & 1.2572 & 0.0267 \\
				\hline
			\end{tabular}
		}
		\hspace*{-2cm}
	}
\end{table}

We implemented the updatable encryption scheme based on $\tx{FrodoPKE}$ in the Python programming language using the NumPy library. Our implementation is based on the official FrodoKEM reference implementation. While Python is not typically considered a high-performance language, the main objective here is to provide a comparative analysis of the execution time for individual components of the scheme. Python remains a popular choice for such tasks due to its simplicity and the availability of powerful scientific libraries. In particular, NumPy offers efficient tools for handling numerical computations, including large matrix operations.

We conducted 100 independent test runs to measure execution times, analyzing how they vary across algorithm components, parameter sets, and input randomness. Among all components, decryption consistently showed the lowest computation time with minimal variance. In contrast, token generation proved to be the most computationally demanding step, primarily due to the multiplication of large matrices whose dimensions grow with the security level of the chosen $\tx{FrodoPKE}$ variant. This results in a significant increase in runtime as the security level rises. However, this performance cost is less critical in practice, as token generation is performed by a trusted party and occurs outside the main ciphertext update process. Moreover, tokens are typically generated infrequently - only once per key rotation - and the lifespan of a key in an updatable encryption scheme is generally measured in days or longer. As a result, the high computational cost of token generation does not hinder the practical deployment of the scheme. These findings illustrate the expected trade-offs between post-quantum security levels and computational performance, particularly in the context of token generation and update operations.
\nocite{*}
\bibliographystyle{fundam}
\bibliography{citations}

\begin{thebibliography}{10}
\providecommand{\url}[1]{\texttt{#1}}
\providecommand{\urlprefix}{URL }
\expandafter\ifx\csname urlstyle\endcsname\relax
  \providecommand{\doi}[1]{doi:\discretionary{}{}{}#1}\else
  \providecommand{\doi}{doi:\discretionary{}{}{}\begingroup
  \urlstyle{rm}\Url}\fi
\providecommand{\eprint}[2][]{\url{#2}}

\bibitem{Frodo}
Alkim E, Avanzi R, Bos J, Ducas L, de~la Piedra A, P{\"o}ppelmann T, Schwabe P,
  Stebila D.
\newblock {FrodoKEM Learning With Errors Key Encapsulation, Algorithm
  Specifications and Supporting Documentation}.
\newblock 2021.
\newblock pp. 1--59.

\bibitem{Frodo2}
Alkim E, Avanzi R, Bos J, Ducas L, de~la Piedra A, P{\"o}ppelmann T, Schwabe P,
  Stebila D.
\newblock {FrodoKEM: Learning With Errors Key Encapsulation. Preliminary
  Standardization Proposal (submitted to ISO)}.
\newblock 2023.
\newblock pp. 1--59.

\bibitem{asharov2012multiparty}
Asharov G, Jain A, L{\'o}pez-Alt A, Tromer E, Vaikuntanathan V, Wichs D.
\newblock Multiparty computation with low communication, computation and
  interaction via threshold FHE.
\newblock In: Advances in Cryptology--EUROCRYPT 2012: 31st Annual International
  Conference on the Theory and Applications of Cryptographic Techniques,
  Cambridge, UK, April 15-19, 2012. Proceedings 31. Springer, 2012 pp.
  483--501.

\bibitem{BEKS20}
Boneh D, Eskandarian S, Kim S, Shih M.
\newblock Improving Speed and Security in Updatable Encryption Schemes.
\newblock In: Advances in Cryptology – ASIACRYPT 2020, volume 12491 of
  \emph{Lecture Notes in Computer Science}. Springer, 2020 pp. 559--589.
\newblock \doi{10.1007/978-3-030-64837-4_19}.

\bibitem{boneh2013key}
Boneh D, Lewi K, Montgomery H, Raghunathan A.
\newblock {Key Homomorphic PRFs and Their Applications}.
\newblock In: Annual Cryptology Conference. Springer, 2013 pp. 410--428.

\bibitem{BDGJ20}
Boyd C, Davies GT, Gjøsteen K, Jiang Y.
\newblock Fast and Secure Updatable Encryption.
\newblock In: Advances in Cryptology – CRYPTO 2020, volume 12171 of
  \emph{Lecture Notes in Computer Science}. Springer, 2020 pp. 464--493.
\newblock \doi{10.1007/978-3-030-56880-1_16}.

\bibitem{brakerski2014leveled}
Brakerski Z, Gentry C, Vaikuntanathan V.
\newblock {(Leveled) Fully Homomorphic Encryption without Bootstrapping}.
\newblock \emph{ACM Transactions on Computation Theory (TOCT)}, 2014.
\newblock \textbf{6}(3):1--36.

\bibitem{brakerski2011fully}
Brakerski Z, Vaikuntanathan V.
\newblock {Fully Homomorphic Encryption from Ring-LWE and Security for Key
  Dependent Messages}.
\newblock In: Annual cryptology conference. Springer, 2011 pp. 505--524.

\bibitem{brakerski2014efficient}
Brakerski Z, Vaikuntanathan V.
\newblock Efficient fully homomorphic encryption from (standard) LWE.
\newblock \emph{SIAM Journal on computing}, 2014.
\newblock \textbf{43}(2):831--871.

\bibitem{cash2012bonsai}
Cash D, Hofheinz D, Kiltz E, Peikert C.
\newblock {Bonsai Trees, or How to Delegate a Lattice Basis}.
\newblock \emph{{Journal of Cryptology}}, 2012.
\newblock \textbf{25}:601--639.

\bibitem{CLT20}
Chen L, Li Y, Tang Q.
\newblock CCA Updatable Encryption Against Malicious Re-Encryption Attacks.
\newblock In: Advances in Cryptology – ASIACRYPT 2020, volume 12491 of
  \emph{Lecture Notes in Computer Science}. Springer, 2020 pp. 590--620.
\newblock \doi{10.1007/978-3-030-64837-4_20}.

\bibitem{EPRS17}
Everspaugh A, Paterson KG, Ristenpart T, Scott S.
\newblock Key Rotation for Authenticated Encryption.
\newblock In: Advances in Cryptology – CRYPTO 2017, volume 10401 of
  \emph{Lecture Notes in Computer Science}. Springer, 2017 pp. 98--129.
\newblock \doi{10.1007/978-3-319-63688-7_4}.

\bibitem{gentry2008trapdoors}
Gentry C, Peikert C, Vaikuntanathan V.
\newblock {Trapdoors for Hard Lattices and New Cryptographic Constructions}.
\newblock In: Proceedings of the fortieth annual ACM symposium on Theory of
  computing. 2008 pp. 197--206.

\bibitem{goldreich2001foundations}
Goldreich O.
\newblock {Foundations of Cryptography: Volume 1, Basic Tools}.
\newblock Cambridge University Press, 2003.

\bibitem{jiang2020}
Jiang Y.
\newblock {The Direction of Updatable Encryption does not Matter Much}.
\newblock In: Advances in Cryptology--ASIACRYPT 2020: 26th International
  Conference on the Theory and Application of Cryptology and Information
  Security, Daejeon, South Korea, December 7--11, 2020, Proceedings, Part III
  26. Springer, 2020 pp. 529--558.

\bibitem{jiang2023backward}
Jiang-Galteland Y, Pan J.
\newblock Backward-Leak Uni-Directional Updatable Encryption from (Homomorphic)
  Public Key Encryption.
\newblock In: Canteaut A, Ishai Y (eds.), Public-Key Cryptography -- PKC 2023,
  volume 13940 of \emph{Lecture Notes in Computer Science}. Springer, 2023 pp.
  399--428.
\newblock \doi{10.1007/978-3-031-31368-4_14}.

\bibitem{klooss2019r}
Kloo{\ss} M, Lehmann A, Rupp A.
\newblock {(R) CCA Secure Updatable Encryption with Integrity Protection}.
\newblock In: Advances in Cryptology--EUROCRYPT 2019: 38th Annual International
  Conference on the Theory and Applications of Cryptographic Techniques,
  Darmstadt, Germany, May 19--23, 2019, Proceedings, Part I 38. Springer, 2019
  pp. 68--99.

\bibitem{LT18}
Lehmann A, Tackmann B.
\newblock Updatable Encryption with Post-Compromise Security.
\newblock In: Advances in Cryptology – EUROCRYPT 2018, volume 10821 of
  \emph{Lecture Notes in Computer Science}. Springer, 2018 pp. 685--716.
\newblock \doi{10.1007/978-3-319-78375-8_22}.

\bibitem{micciancio2002complexity}
Micciancio D, Goldwasser S.
\newblock {Complexity of Lattice Problems: a Cryptographic Perspective}, volume
  671.
\newblock Springer Science \& Business Media, 2002.

\bibitem{micciancio2007worst}
Micciancio D, Regev O.
\newblock {Worst-Case to Average-Case Reductions Based on Gaussian Measures}.
\newblock \emph{{SIAM Journal on Computing}}, 2007.
\newblock \textbf{37}(1):267--302.

\bibitem{nishimaki2022direction}
Nishimaki R.
\newblock The direction of updatable encryption does matter.
\newblock In: IACR International Conference on Public-Key Cryptography.
  Springer, 2022 pp. 194--224.

\bibitem{peikert2015decade}
Peikert C.
\newblock {A Decade of Lattice Cryptography}.
\newblock \emph{{Cryptology ePrint Archive}}, 2015.

\bibitem{peikert2006efficient}
Peikert C, Rosen A.
\newblock {Efficient Collision-Resistant Hashing from Worst-Case Assumptions on
  Cyclic Lattices}.
\newblock In: {Theory of Cryptography: Third Theory of Cryptography Conference,
  TCC 2006, New York, NY, USA, March 4-7, 2006. Proceedings 3}. Springer, 2006
  pp. 145--166.

\bibitem{regev2009lattices}
Regev O.
\newblock {On Lattices, Learning with Errors, Random Linear Codes, and
  Cryptography}.
\newblock \emph{{Journal of the ACM (JACM)}}, 2009.
\newblock \textbf{56}(6):1--40.

\bibitem{SS21}
Slamanig D, Striecks C.
\newblock Revisiting Updatable Encryption: Controlled Forward Security,
  Constructions and a Puncturable Perspective.
\newblock Cryptology ePrint Archive, Paper 2021/268, 2023.
\newblock \url{https://eprint.iacr.org/2021/268/20231003:110230}.

\end{thebibliography}

%%%%%%%%%%%%%%%%%%%%%%%%%%%%%%%%%%%%%%%%%%%%%%%%%%%%%%%%%%%%%%%%%%%%%%

\end{document}